\documentclass[12pt]{article}
\usepackage{verbatim,graphics,graphicx,color,slashed,amsmath}
\usepackage{bm}
\usepackage{cite}
\usepackage{amssymb}
\usepackage{braket}
\usepackage{ascmac}
\usepackage{multirow}
\usepackage[compat=1.1.0]{tikz-feynman} 
\usepackage[]{hyperref}
\hypersetup{colorlinks,bookmarks,unicode,linktocpage=true,linkcolor=blue, anchorcolor=blue, citecolor=blue}
\newcommand{\be}{\begin{equation}}
\newcommand{\ee}{\end{equation}}
\newcommand{\bea}{\begin{eqnarray}}
\newcommand{\eea}{\end{eqnarray}}

\newcounter{num}

\begin{document}

\begin{titlepage}
%
%


%

\begin{centering}
\vspace{1cm}
{\Large {\bf Positivity Bounds on Higgs-Portal Dark Matter}} \\

\vspace{1.5cm}

{\bf Seong-Sik Kim$^{1,\ddagger}$, Hyun Min Lee$^{1,\dagger}$, and Kimiko Yamashita$^{1,2,\star}$ }
\vspace{.5cm}

{\it  $^1$Department of Physics, Chung-Ang University, Seoul 06974, Korea.}  \\[0.2cm]
{\it $^2$Department of Physics, Ibaraki University, Mito 310-8512, Japan.}

\vspace{.5cm}


\end{centering}
\vspace{2cm}

\begin{abstract}
\noindent
We consider the positivity bounds  for WIMP scalar dark matter with effective Higgs-portal couplings up to dimension-8 operators. Taking the superposed states for Standard Model Higgs and scalar dark matter, we show that the part of the parameter space for the effective couplings, otherwise unconstrained by phenomenological bounds, is ruled out by the positivity bounds on the dimension-8 derivative operators. We find that dark matter relic density, direct and indirect detection and LHC constraints are complementary to the positivity bounds in constraining the effective Higgs-portal couplings. In the effective theory obtained from massive graviton or radion,  there appears a correlation between dimension-8 operators and other effective Higgs-portal couplings for which the strong constraint from direct detection can be evaded.  
Nailing down the parameter space mainly by relic density, direct detection and positivity bounds, we find that there are observable cosmic ray signals coming from the dark matter annihilations into a pair of Higgs bosons, $WW$ or $ZZ$. 
\end{abstract}
\vspace{3cm}

\begin{flushleft} 
${}^{\ddagger}$Email: sskim.working@gmail.com\\
${}^{\dagger}$Email: hminlee@cau.ac.kr\\
${}^{\star}$Email: kimikoy@cau.ac.kr
\end{flushleft}

\end{titlepage}

\baselineskip 18pt

\newpage

\section{Introduction}\label{sec:introduction}

Positivity bounds are the general feature of a local Quantum Field Theory (QFT) embeddable into the Ultra-Violet (UV) complete theory \cite{Adams:2006sv,Pham:1985cr,Ananthanarayan:1994hf}. 
We rely on the fundamental principles in QFT such as special relativity, conservation of probability, and causality, which correspond to Lorentz invariance, unitarity, and analyticity of transition amplitudes, respectively. Then, also using the dispersion relation and the optical theorem, we can obtain positivity bounds for the local operators in Effective Field Theory (EFT).

The best bounds on the dimension-8 operators can be obtained from the forward limit of elastic scattering (i.e. $t=0$),  thanks to the crossing symmetry~\cite{Zhang:2021eeo}.
There are more bounds obtainable in the other approaches using the superposition of the states and extremal rays than in the elastic scattering approaches~\cite{Bi:2019phv,Remmen:2019cyz,Ghosh:2022qqq,Zhang:2020jyn,Yamashita:2020gtt,Zhang:2021eeo,Li:2022aby}. It is remarkable that the extremal ray approach may give a chance to expose other bounds, enabling us to connect the region of the Wilson
coefficient space bounded by positivity to the UV physics. 

In this article, considering the effective theory for the Higgs fields in the Standard Model (SM) and scalar dark matter with $Z_2$ parity, we derive the positivity bounds on the dimension-8 derivative Higgs-portal couplings from the forward $2\to 2$ elastic scattering amplitudes. To this purpose, it is sufficient for us to take  the superposed states for Higgs and scalar dark matter, because it does not rely on a precise form of the UV physics.
On top of the positivity bounds, we also impose various phenomenological constraints from dark matter relic density, direct and indirect detection for dark matter, and show the interplay of them with the positivity bounds in constraining the effective Higgs-portal couplings. 
We discuss the positivity bounds in relation to the effective theory obtained after massive graviton and/or radion are integrated out.

The paper is organized as follows. 
First, we introduce the general effective interactions for Higgs and scalar dark matter up to dimension-8 operators and show some benchmark UV models  where  a set of correlated effective interactions is obtained after massive graviton or radion are integrated out. Next, we show the positivity bounds on the dimension-8 Higgs-portal interactions based on the elastic scattering for the superposed states. Then, we consider the dark matter relic density, direct and indirect detection, and Large Hadron Collider (LHC) constraints on the effective Higgs-portal interactions. Finally, conclusions are drawn.  There is one appendix dealing with the one-loop corrections to the Higgs-portal dimension-8 operators in the presence of dimension-6 operators.

\section{Higgs-portal couplings in EFT}\label{sec:eft}

We take the effective Lagrangian for the real scalar dark matter $\varphi$ and the Higgs doublet  $H$, up to dimension-8 operators, as
\bea
{\cal L}_{\rm Higgs-portal}= {\cal L}_1 +{\cal L}_2
\eea
with
\bea
{\cal L}_1 &=&-\frac{1}{6\Lambda^4} \Big(c_1 m_\varphi^4 \varphi^4 + 4c_2 m^4_H |H|^4 + 8c'_2\lambda_H  m^2_H |H|^6 + 4c^{\prime\prime}_2\lambda^2_H  |H|^8 \nonumber\\
&&\quad+ 4 c_3 m^2_\varphi m^2_H \varphi^2 |H|^2+4c'_3 \lambda_H  m^2_\varphi \varphi^2 |H|^4  \Big) \nonumber \\
&&+\frac{1}{6\Lambda^4} \bigg( d_1 m^2_\varphi \varphi^2 (\partial_\mu\varphi)^2 + 4d_2 m^2_H |H|^2 |D_\mu H|^2+ 4d'_2\lambda_H  |H|^4 |D_\mu H|^2  \nonumber \\
&&\quad+2d_3 m_\varphi^2 \varphi^2 |D_\mu H|^2+ 2d_4 m^2_H |H|^2 (\partial_\mu\varphi)^2+ 2d'_4\lambda_H  |H|^4 (\partial_\mu\varphi)^2 \bigg),  \label{lag1} \\
{\cal L}_2&=& \frac{C^{(1)}_{H^2\varphi^2}}{\Lambda^4} O^{(1)}_{H^2\varphi^2}+ \frac{C^{(2)}_{H^2\varphi^2}}{\Lambda^4} O^{(2)}_{H^2\varphi^2} \nonumber \\
&&+\frac{ C_{\varphi^4}}{\Lambda^4} O_{\varphi^4}+\frac{C^{(1)}_{H^4}}{\Lambda^4} O^{(1)}_{H^4}+\frac{C^{(2)}_{H^4}}{\Lambda^4} O^{(2)}_{H^4} +\frac{C^{(3)}_{H^4}}{\Lambda^4} O^{(3)}_{H^4} \label{effdim8}
\eea
where $c_{1,2,3}, d_{1,2,3,4}$ and primed quantities are dimensionless parameters, and  $C^{(1)}_{H^2\varphi^2}, C^{(2)}_{H^2\varphi^2}, C_{\varphi^4}$ and $C^{(1,2,3)}_{H^4}$  are the Wilson coefficients for the dimension-8 operators containing four derivatives listed in  Table \ref{tab:dim8_operators}, and $\Lambda$ is the cutoff scale. 

\begin{table}[hbt!]
\center
\begin{tabular}{| l  l |}
\hline
$O^{(1)}_{H^2\varphi^2} = (D_\mu H^\dagger D_\nu H)(\partial^\mu \varphi \partial^\nu \varphi)$
& $O^{(2)}_{H^2\varphi^2} = (D_\mu H^\dagger D^\mu H)(\partial_\nu \varphi \partial^\nu \varphi)$\\
\hline\hline
$O_{\varphi^4} = \partial_\mu \varphi \partial^\mu \varphi \partial_\nu \varphi \partial^\nu \varphi$ & \\
\hline\hline
$O^{(1)}_{H^4} = (D_\mu H^\dagger D_\nu H)(D^\nu H^\dagger D^\mu H)$
& $O^{(2)}_{H^4} = (D_\mu H^\dagger D_\nu H)(D^\mu H^\dagger D^\nu H)$ \\
$O^{(3)}_{H^4} = (D_\mu H^\dagger D^\mu H)(D_\nu H^\dagger D^\nu H)$&\\
\hline
\end{tabular}
\caption{Dimension-8 operators for Higgs and real scalar dark matter}
\label{tab:dim8_operators}
\end{table}

We assume the $Z_2$ symmetry for the scalar dark matter, so the effective Higgs-portal interactions include only even numbers of scalar dark matter particles.  Moreover, the Higgs mass parameter $m^2_H$ is introduced in the parametrization of the effective Higgs-portal couplings, so it can be rewritten as $m^2_H=-\lambda_H v^2$ after electroweak symmetry is broken dominantly for the renormalizable Higgs potential.

After electroweak symmetry breaking with $H=(0,v)^T/\sqrt{2}$ and using $m^2_H=-\lambda_H v^2$, the effective Higgs-portal interactions linear in the Higgs boson $h$ are given by
\bea
{\cal L}_{h,{\rm linear}} =\frac{1}{3\Lambda^4}\,h \bigg[2(c_3-c'_3) \lambda_H v^3 m^2_\varphi \varphi^2-(d_4-d'_4) \lambda_H v^3 (\partial_\mu\varphi)^2  \bigg], \label{Higgslinear}
\eea
which vanishes for $c'_3=c_3$ and $d'_4=d_4$. Then, there is no effective linear Higgs-portal coupling to scalar dark matter, so there is no tree-level contribution to dark matter annihilation or DM-nucleon scattering processes with Higgs exchanges, being consistent with the results for the massive spin-2 particle in Ref.~\cite{GDD}.  The origin of the effective Higgs-portal couplings will be discussed shortly. 
Henceforth, we also use the physical Higgs mass, $m^2_h=2\lambda_H v^2=-2m^2_H$, alternatively.

\subsection{Graviton-like interactions}

The dimension-8 operators as well as the lower dimensional operators in Eqs.~(\ref{lag1}) and (\ref{effdim8}) can be originated from the exchanges of a massive spin-2 particle $G_{\mu\nu}$ \cite{Lee:2013bua}. Introducing the interactions between a massive spin-2 particle and Higgs/dark matter by the energy-momentum tensors,
\bea
{\cal L}_{G} = -\frac{c_H}{M} \, G^{\mu\nu} T^H_{\mu\nu} - \frac{c_\varphi}{M}\, G^{\mu\nu} T^\varphi_{\mu\nu}
\eea
where $T^H_{\mu\nu},  T^\varphi_{\mu\nu}$ are the energy-momentum tensors for Higgs and dark matter, given by
\bea
T^H_{\mu\nu} &=&(D_\mu H)^\dagger D_\nu H + (D_\nu H)^\dagger D_\mu H   - g_{\mu\nu}  [g^{\rho\sigma} (D_\rho H)^\dagger D_\sigma H] \nonumber \\
&&+ g_{\mu\nu} (m_H^2 |H|^2+\lambda_H |H|^4), \\
T^\varphi_{\mu\nu} &=& \partial_\mu \varphi \partial_\nu\varphi  -\frac{1}{2} g_{\mu\nu} (g^{\rho\sigma} \partial_\rho \varphi\partial_\sigma \varphi) +\frac{1}{2} g_{\mu\nu} m_\varphi^2 \varphi^2,
\eea
and $c_H, c_\varphi$ are dimensionless parameters, and $M$ is the suppression scale for the spin-2 interactions. Then, after integrating out the massive spin-2 particle, we obtain the effective Lagrangian for Higgs and scalar dark matter \cite{GDD}, as follows,
\bea
{\cal L}_{G,{\rm eff}}&=&\frac{1}{2m^2_GM^2}\, T^{\mu\nu} P_{\mu\nu,\alpha\beta} T^{\alpha\beta}  \nonumber \\
 &=&\frac{1}{4m^2_GM^2} \bigg( 2 T_{\mu\nu}T^{\mu\nu}-\frac{2}{3} T^2 \bigg),
\eea
where $T_{\mu\nu}=c_H T^H_{\mu\nu}+c_\varphi T^\varphi_{\mu\nu}$, $T=c_H T^H+c_\varphi T^\varphi$ with $T^H=T^{H,\mu}_\mu$ , $T^\varphi=T^{\varphi,\mu}_\mu$, and the polarization tensor for the massive gravition is given by $P_{\mu\nu,\alpha\beta}=\frac{1}{2}\big(g_{\mu\alpha}g_{\nu\beta}+g_{\mu\beta}g_{\nu\alpha}-\frac{2}{3} g_{\mu\nu}g_{\alpha\beta}\big)$.  Then, we get the effective dimension-8 operators with the Wilson coefficients,
\bea
\frac{C^{(2)}_{H^2\varphi^2}}{\Lambda^4}=-\frac{1}{3} \frac{C^{(1)}_{H^2\varphi^2}}{\Lambda^4}=-\frac{2c_H c_\varphi}{3m^2_G M^2}, \label{G1}
\eea
and the effective lower dimensional Higgs-portal operators with
\bea
\frac{c^{(\prime)}_3}{\Lambda^4}=\frac{d_3}{\Lambda^4}=\frac{d^{(\prime)}_4}{\Lambda^4}= \frac{c_H c_\varphi}{m^2_G M^2}=\frac{1}{2}\frac{C^{(1)}_{H^2\varphi^2}}{\Lambda^4}.  \label{G2}
\eea

Similarly, we can also match the effective self-interactions for Higgs and scalar dark matter due to the spin-2 exchanges, as follows,
\bea
\frac{C_{\varphi^4}}{\Lambda^4}&=&\frac{c^2_\varphi}{3m^2_G M^2}= \frac{c_\varphi}{6c_H} \frac{C^{(1)}_{H^2\varphi^2}}{\Lambda^4}, \label{G3} \\
 \frac{C^{(1)}_{H^4}}{\Lambda^4}&=& \frac{C^{(2)}_{H^4}}{\Lambda^4}= -\frac{3}{2}\frac{C^{(3)}_{H^4}}{\Lambda^4}=\frac{c^2_H}{m^2_G M^2}=\frac{c_H}{2c_\varphi}\frac{C^{(1)}_{H^2\varphi^2}}{\Lambda^4}, \label{G4} \\
\frac{c_1}{\Lambda^4}&=&\frac{d_1}{\Lambda^4}= \frac{c^2_\varphi}{m^2_G M^2}=\frac{c_\varphi}{2c_H}\,\frac{C^{(1)}_{H^2\varphi^2}}{\Lambda^4}, \label{G5} \\ 
\frac{c^{(\prime,\prime\prime)}_2}{\Lambda^4}&=&\frac{d^{(\prime)}_2}{\Lambda^4}= \frac{c^2_H}{m^2_G M^2}=\frac{c_H}{2c_\varphi} \,\frac{C^{(1)}_{H^2\varphi^2}}{\Lambda^4}. \label{G6}
\eea
In this case, we can find the correlations between the effective couplings for Higgs and scalar dark matter up to dimension-8 operators from the tree-level matching conditions in terms of a single effective portal-coupling, $C^{(1)}_{H^2\varphi^2}$, whereas the Higgs and scalar self-interactions are subject to one more parameter, $\frac{c_\varphi}{c_H}$, originated from a more fundamental theory.

Consequently, for $c_H c_\varphi>0$, which is the case for the attractive force between Higgs and dark matter, we obtain $ C^{(1)}_{H^2\varphi^2}=-3 C^{(2)}_{H^2\varphi^2}>0$ for the dimension-8 Higgs-portal couplings at the matching scale of the massive spin-2 particle.

\subsection{Radion-like interactions}

We now consider another way to match the effective Higgs-portal interactions in the presence of a radion-like scalar particle $r$ \cite{Lee:2013bua}. We introduce the interactions between the radion from the extra dimension (or dilaton) and Higgs/matter by the trace of the energy-momentum tensors,
\bea
{\cal L}_r= \frac{c^r_H}{\sqrt{6}M} r\, T^H +\frac{c^r_\varphi}{\sqrt{6}M} r\, T^\varphi
\eea
with radion couplings, $c^r_H, c^r_\varphi$.
Then, integrating out the dilaton-like scalar $r$, we obtain the effective Lagrangian in the following form,
\bea
{\cal L}_{r,{\rm eff}}=\frac{1}{12m^2_r M^2}\, T^2
\eea
with $T=c^r_H T^H+c^r_\varphi T^\varphi$.
Therefore, the Wilson coefficients of the  resultant effective dimension-8 operators become
\bea
C^{(1)}_{H^2\varphi^2} =0, \quad \frac{C^{(2)}_{H^2\varphi^2}}{\Lambda^4} = \frac{c^r_H c^r_\varphi}{3m^2_r M^2}, \label{r1}
\eea
and the effective lower dimensional Higgs-portal operators are given by
\bea
\frac{c^{(\prime)}_3}{\Lambda^4}=\frac{d_3}{\Lambda^4}=\frac{d^{(\prime)}_4}{\Lambda^4}= -\frac{2c^r_H c^r_\varphi}{m^2_r M^2}=-6\frac{C^{(2)}_{H^2\varphi^2}}{\Lambda^4}. \label{r2}
\eea
Thus, as for the massive spin-2 particle, for $c'_3=c_3$ and $d'_4=d_4$, there is no effective linear Higgs-portal coupling to scalar dark matter.

Similarly, the effective self-interactions for Higgs and dark matter are given by
\bea
\frac{C_{\varphi^4}}{\Lambda^4}&=&\frac{(c^r_\varphi)^2}{12m^2_r M^2}= \frac{c^r_\varphi}{4c^r_H} \frac{C^{(2)}_{H^2\varphi^2}}{\Lambda^4}, \label{r3}\\
C^{(1)}_{H^4}&=&C^{(2)}_{H^4}=0,\quad  \frac{C^{(3)}_{H^4}}{\Lambda^4}=\frac{(c^r_H)^2}{3m^2_r M^2}=\frac{c^r_H}{c^r_\varphi}\frac{C^{(2)}_{H^2\varphi^2}}{\Lambda^4}, \label{r4} \\
\frac{c_1}{\Lambda^4}&=&\frac{d_1}{\Lambda^4}= -\frac{2(c^r_\varphi)^2}{m^2_r M^2}=-\frac{6c^r_\varphi}{c^r_H}\,\frac{C^{(2)}_{H^2\varphi^2}}{\Lambda^4}, \label{r5} \\ 
\frac{c^{(\prime,\prime\prime)}_2}{\Lambda^4}&=&\frac{d^{(\prime)}_2}{\Lambda^4}= -\frac{2(c^r_H)^2}{m^2_r M^2}=-\frac{6c^r_H}{c^r_\varphi} \,\frac{C^{(2)}_{H^2\varphi^2}}{\Lambda^4}. \label{r6}
\eea

Consequently, for $c^r_H c^r_\varphi>0$, which is the case for the attractive force between Higgs and dark matter due to the radion, we obtain $ C^{(2)}_{H^2\varphi^2}>0$ for the dimension-8 Higgs-portal couplings at the matching scale of the radion. 

Summing up both massive spin-2 and radion couplings in Eqs.~(\ref{G1})-(\ref{G6}) and (\ref{r1})-(\ref{r6}), we get the effective Higgs-portal interactions,
\bea
\frac{C^{(1)}_{H^2\varphi^2}}{\Lambda^4}&=&\frac{2c_Hc_\varphi}{m^2_G M^2}, \quad  \frac{C^{(2)}_{H^2\varphi^2}}{\Lambda^4} =-\frac{2c_H c_\varphi}{3m^2_GM^2}+ \frac{c^r_H c^r_\varphi}{3m^2_r M^2}, \label{t1} \\
\frac{c^{(\prime)}_3}{\Lambda^4}&=&\frac{d_3}{\Lambda^4}=\frac{d^{(\prime)}_4}{\Lambda^4}=\frac{c_Hc_\varphi}{m^2_GM^2} -\frac{2c^r_H c^r_\varphi}{m^2_r M^2}=-\frac{3}{2} \frac{C^{(1)}_{H^2\varphi^2}}{\Lambda^4}-6\frac{C^{(2)}_{H^2\varphi^2}}{\Lambda^4},  \label{t2} 
\eea
and the effective self-interactions, 
\bea
\frac{C_{\varphi^4}}{\Lambda^4}&=&\frac{c^2_\varphi}{3m^2_G M^2}+\frac{(c^r_\varphi)^2}{12m^2_r M^2}, \label{t3}\\
\frac{C^{(1)}_{H^4}}{\Lambda^4}&=&\frac{C^{(2)}_{H^4}}{\Lambda^4}=\frac{c^2_H}{m^2_G M^2},\quad  \frac{C^{(3)}_{H^4}}{\Lambda^4}=-\frac{2c^2_H}{3m^2_G M^2}+\frac{(c^r_H)^2}{3m^2_r M^2}, \label{t4} \\
\frac{c_1}{\Lambda^4}&=&\frac{d_1}{\Lambda^4}= \frac{c^2_\varphi}{m^2_GM^2}-\frac{2(c^r_\varphi)^2}{m^2_r M^2}, \label{t5} \\ 
\frac{c^{(\prime,\prime\prime)}_2}{\Lambda^4}&=&\frac{d^{(\prime)}_2}{\Lambda^4}=\frac{c^2_H}{m^2_GM^2} -\frac{2(c^r_H)^2}{m^2_r M^2}=-3\frac{C^{(1)}_{H^4}}{\Lambda^4}-6\frac{C^{(3)}_{H^4}}{\Lambda^4}. \label{t6}
\eea
Here, we note that only the massive spin-2 particle contributes to $C^{(1)}_{H^2\varphi^2}$  while both massive spin-2 particle and radion contribute to $C^{(2)}_{H^2\varphi^2}$. On the other hand, the dimension-4 and dimension-6 Higgs-portal couplings are universally determined in terms of $C^{(1)}_{H^2\varphi^2}$ and $C^{(2)}_{H^2\varphi^2}$ in Eq.~(\ref{t2}), so there is no tree-level contribution to the DM-nucleon scattering, as discussed below Eq.~(\ref{Higgslinear}) in the beginning of this section.  Moreover, the dimension-4 Higgs portal coupling, $\varphi^2|H|^2$, is doubly suppressed by $\frac{m^2_\varphi}{\Lambda^2}$ and $\frac{m^2_H}{\Lambda^2}$. 

%
\section{Positivity bounds on dimension-8 Higgs-portal}\label{sec:positive}

In this article, we derive the positivity bounds on the dimension-8 derivative Higgs-portal couplings for scalar dark matter.

\subsection{Scattering amplitudes for superposed states}

Briefly speaking, the statement of the positivity bounds is that the second order $s$-derivative of the amplitudes with poles subtracted must be positive.
For the case of the superposed elastic scattering process, $ab \to ab$, 
where $a$ and $b$ states are superposition states with coefficients characterized by $u$ and $v$ vectors,
\begin{align}
\ket{a} = u^i \ket{i}, \ket{b} = v^i \ket{i},
\end{align}
we get the positivity bounds by
\begin{align}
u^i v^j u^{*k}  v^{*l}  M^{ijkl} \geq 0,
\end{align}
where $i, j, k, l$ indices run through the number of the states involved in the superposition and 
\begin{align}
M^{ijkl} = \frac{1}{2} \frac{d^2}{ds^2} M (i j \to k l)(s, t=0)\bigg|_{s\to 0}.
\end{align}
Here, we assumed that the low-energy poles are subtracted in the scattering amplitude $M(ij \to kl)$.

In the case of a real scalar dark matter with Higgs-portal couplings, $a$ and $b$ states in $ab \to ab$ scattering, correspond to
\bea
\ket{a} &=& \sum^{4}_{i=1} u_i \ket{\phi_i} + u_5 \ket{\varphi},\\
\ket{b} &=& \sum^{4}_{i=1} v_i \ket{\phi_i} + v_5 \ket{\varphi},
\eea
respectively. Here, $\varphi$ is the real scalar dark matter field and 
the SM Higgs doublet  $H$ is written in terms of four real scalar fields, $\phi_i (i=1,2,3,4)$,  as follows,
\begin{align}
H =\frac{1}{\sqrt{2}}\begin{pmatrix}
\phi_1 + i\phi_2 \\
\phi_3 + i\phi_4
\end{pmatrix}.
\end{align}

The effective operators involved in this superposed elastic scattering process are not only Higgs-portal operators for real scalar dark matter given in the first line in Table~\ref{tab:dim8_operators}, but also
operators including the dark matter only (in the second line) and the Higgs doublet only (in the third and fourth lines), respectively.
The full operators involved in this superposed elastic scattering process contributing on the positivity bounds are listed in Table~\ref{tab:dim8_operators}.

Performing the calculations of all the amplitudes involved in the processes by
FeynRules, FeynArts and FormCalc~\cite{Christensen:2008py,Alloul:2013bka,Hahn:2000kx}, 
we find the amplitude for the superposed states,
\begin{align}
M \equiv u_i v_j u_{*k}  v_{*l} M^{ijkl} &= (X_1 + Y_1 + Z_1) C^{(2)}_{H^4} + Y_1 C^{(3)}_{H^4} + (X_1 + Y_1)C^{(1)}_{H^4}& \nonumber \\
&\, \, \, \, + (X_2 + Z_2) C^{(1)}_{H^2\varphi^2} +2Z_2 C^{(2)}_{H^2\varphi^2} 
+ 4 Y_2 C_{\varphi^4}  \geq 0& \label{eq:amp}
\end{align}
where 
\begin{align}
X_1 &= \frac{1}{2} (-u_4v_1 - u_3v_2 + u_2v_3 + u_1 v_4)^2 + \frac{1}{2} (-u_3v_1 + u_4v_2 + u_1v_3 - u_2 v_4)^2, \label{eq:x1}\\
Y_1 &= (u_1v_1 + u_2v_2 + u_3v_3 + u_4v_4)^2,\\
Z_1 &= (u_2v_1 - u_1v_2 + u_4v_3 - u_3v_4)^2, \label{eq:z1}\\
X_2 &=  \frac{1}{2} \left(u^2_5(v^2_1 + v^2_2 + v^2_3 + v^2_4)+ v^2_5(u^2_1 + u^2_2 + u^2_3 + u^2_4)\right), \\
Y_2 &= (u_5 v_5)^2,  \label{eq:y2}\\
Z_2 &= u_5v_5(u_1v_1 + u_1v_2 + u_3v_3 + u_4v_4).  \label{eq:z2}
\end{align}
Here, we took the same combinations for $X_1$, $Y_1$, and $Z_1$ as in Ref.~\cite{Yamashita:2020gtt}, and the coefficients $u_i$ and $v_i$ are assumed to be real numbers for simplicity, but taking them to be complex numbers does not give rise to additional constraints,
thanks to the crossing symmetry of the forward scattering amplitudes~\cite{Yamashita:2020gtt}.

\subsection{Positivity bounds}

In order to derive the positivity bounds, we now rewrite the amplitude in Eq.~\eqref{eq:amp} as
\begin{align}
M &= M_1 X_1 + M_2 Y_1 + M_3 Z_1 + M_4 X_2 + M_5 Y_2 + M_6 Z_2 \geq 0 \label{eq: M}
\end{align}
where 
\begin{align}
M_1 = C^{(1)}_{H^4}  + C^{(2)}_{H^4}, \quad M_2 = C^{(1)}_{H^4}  + C^{(2)}_{H^4} + C^{(3)}_{H^4}, \quad M_3 = C^{(2)}_{H^4} , \nonumber \\
M_4 = C^{(1)}_{H^2\varphi^2}, \quad M_5 = 4C_{\varphi^4}, \quad M_6 = C^{(1)}_{H^2\varphi^2} + 2C^{(2)}_{H^2\varphi^2}.  \label{eq:coeffs}
\end{align}
As $X_i$, $Y_i$, and $Z_i$ $(i = 1, 2)$ are quartic polynomials of $u_j$ and $v_j$ $(j = 1-5)$ in Eqs.~(\ref{eq:x1})-(\ref{eq:z2}),  we find the ranges for them under which Eq.~\eqref{eq: M} is positive semidefinite, as follows,
\bea
&&X_1 \geq 0, \quad Y_1 \geq 0, \quad Z_1 \geq 0, \quad Y_2 \geq 0, \quad Z_2 = \pm \sqrt{Y_1 Y_2}, \label{eq:range1} \\
&&X_2 \geq \sqrt{Y_2(2X_1 + Y_1 + Z_1)}. \label{eq:range2} 
\eea
Here, we note that the bounds in Eq.~\eqref{eq:range1} are obtained simply from their definitions in Eqs.~\eqref{eq:x1}--\eqref{eq:z1} and Eqs.~\eqref{eq:y2}--\eqref{eq:z2}, and
Eq.~\eqref{eq:range2} can be derived from
\begin{align}
X^2_2 -Y_2(2X_1 + Y_1 + Z_1) = \frac{1}{4}\left(u^2_5(v^2_1 + v^2_2 + v^2_3 + v^2_4) - v^2_5(u^2_1 + u^2_2 + u^2_3 + u^2_4)\right)^2 \geq 0.
\end{align}

First of all, taking $M \geq 0$ from the positivity argument, we immediately obtain
\begin{align}
M_1 \geq 0, \quad M_2 \geq 0, \quad M_3\geq 0, \quad M_4 \geq 0, \quad M_5 \geq 0. \label{eq:bound1}
\end{align}
For example, we obtain $M_1 \geq 0$ by setting $X_1 \neq 0$ and others to zero by $u_1 =1$, $v_3=1$ but other coefficients set to zero. Then,  this corresponds to the positivity bound for the elastic forward scattering with $\ket{\phi_1}$ and $\ket{\phi_3}$.

For deriving the other bounds, we should minimize Eq.~\eqref{eq: M} in the ranges of Eqs.~\eqref{eq:range1} and \eqref{eq:range2}.
When $Y_1 =0$, we just obtain a trivial result, $M \geq 0$.
So, we concentrate on a nonzero $Y_1$, i.e., $Y_1 > 0$ from Eq.~\eqref{eq:range1}. Then, the problem becomes to minimize
\bea
X_1 M_1 + Y_1 M_2 + Z_1 M_3 + X_2 M_4 + Y_2 M_5 + Z_2 M_6 \, \, (\geq 0),
\label{eq:positivity_d_1}
\eea
subject to
\bea
X_1 \geq 0, Y_1 > 0, Z_1 \geq 0, X_2 \geq \sqrt{Y_2(2X_1 + Y_1 + Z_1)}, Y_2 \geq 0, Z_2 = \pm \sqrt{Y_1 Y_2}.
\eea

Knowing $M_4 \geq 0$ from Eq.~\eqref{eq:bound1}, we take $X_2 =\sqrt{Y_2(2X_1 + Y_1 + Z_1)}$ to minimize the amplitude $M$.
Moreover, as $|M_6| \geq 0$, we can take $M_6Z_2=-\sqrt{Y_1 Y_2}|M_6|$.
Then, after dividing the amplitude $M$ by $Y_1$,
 the problem \eqref{eq:positivity_d_1} is now to minimize
\bea
f(x,y,z) = x^2 M_1 + y^2 M_5 + z^2 M_3 -y(|M_6| -\sqrt{2x^2 + z^2 + 1} M_4) + M_2 \, \, (\geq 0),
\label{eq:positivity_d_2}
\eea
subject to
\bea
x \geq 0, \quad y \geq 0, \quad z \geq 0,
\eea
where $x = \sqrt{\frac{X_1}{Y_1}}$, $y = \sqrt{\frac{Y_2}{Y_1}}$, and $z = \sqrt{\frac{Z_1}{Y_1}}$.
Since $M_1 \geq 0$, $M_3 \geq 0$, and $M_4 \geq 0$ from Eq.~\eqref{eq:bound1}, we further take $x = z = 0$ and now only have to minimize
\bea
f(0,y,0) = y^2 M_5 - y(|M_6| - M_4) + M_2 \, \, (\geq 0), \label{eq:positivity_d_3}
\eea
for $y \geq 0$.
Since $f(0,y,0)$ has a minimum for $y = \frac{|M_6| - M_4}{2 M_5}$, we obtain
\bea
M_2 - \frac{(|M_6| - M_4)^2}{4 M_5} \geq 0, \quad {\rm for}\,\,  |M_6| \geq M_4,
 \label{eq:bound3-1} 
\eea
or
\bea
M_2 \geq 0, \quad  {\rm for}\,\,  |M_6| \leq M_4,
\eea
from  Eq.~\eqref{eq:bound1}.
Since $M_2 \geq 0$ and  $M_5 \geq 0$  from Eq.~\eqref{eq:bound1}, the condition in Eq.~\eqref{eq:bound3-1} is the same as
\bea
2\sqrt{M_2 M_5} \geq |M_6| - M_4.  \label{eq:bound3-2}
\eea
Here, we omitted $|M_6| \geq M_4$ because Eq.~\eqref{eq:bound3-2} is automatically satisfied for $ |M_6| < M_4$.

Substituting the Wilson coefficients with Eq.~\eqref{eq:coeffs} in the positivity conditions in Eq.~\eqref{eq:bound1},
we finally arrive at
\begin{align}
&C^{(1)}_{H^4}  + C^{(2)}_{H^4} \geq 0,  \\ 
&C^{(1)}_{H^4}  + C^{(2)}_{H^4} + C^{(3)}_{H^4} \geq 0,\\ 
&C^{(2)}_{H^4} \geq 0, \\
&C^{(1)}_{H^2\varphi^2} \geq 0, \label{eq:portal1}\\
&C_{\varphi^4} \geq 0, \\
&4\sqrt{(C^{(1)}_{H^4}  + C^{(2)}_{H^4} + C^{(3)}_{H^4})C_{\varphi^4}}
\geq \left|C^{(1)}_{H^2\varphi^2} + 2C^{(2)}_{H^2\varphi^2}\right| - C^{(1)}_{H^2\varphi^2}. \label{eq:boundf}
\end{align}
The corresponding forward elastic scattering channel for each bound is summarized in Table~\ref{tab:bounds_channels}.
Here, for the final two elastic scattering channels, we can divide the positivity condition in Eq.~\eqref{eq:boundf} into the cases, 
$C^{(1)}_{H^2\varphi^2} + 2C^{(2)}_{H^2\varphi^2} \leq 0$ and $C^{(1)}_{H^2\varphi^2} + 2C^{(2)}_{H^2\varphi^2} \geq 0$.

\begin{table}[t]
\center
\begin{tabular}{|c|c|}
\hline
	Bounds & Channels $(\ket 1+\ket 2 \to \ket 1\ +\ket 2)$
	\\\hline\hline
	$C^{(1)}_{H^4}  + C^{(2)}_{H^4} \ge0$ & $\ket 1 = \ket{\phi_1},\ \ket 2=\ket{\phi_3}$
	\\\hline
	$C^{(1)}_{H^4}  + C^{(2)}_{H^4} + C^{(3)}_{H^4}\ge0$ & $\ket 1 = \ket{\phi_1},\ \ket 2=\ket{\phi_1}$
	\\\hline
	$C^{(2)}_{H^4}\ge0$ & $\ket 1 = \ket{\phi_1},\ \ket 2=\ket{\phi_2}$
	\\\hline
	$C^{(1)}_{H^2\varphi^2}\ge0$ & $\ket 1 = \ket{\phi_1},\ \ket 2=\ket{\varphi}$
	\\\hline
	$C_{\varphi^4} \ge0$ & $\ket 1 = \ket{\varphi},\ \ \ket 2=\ket{\varphi}$
	\\\hline
	$ 2\sqrt{(C^{(1)}_{H^4}  + C^{(2)}_{H^4} + C^{(3)}_{H^4})C_{\varphi^4}}$ 
	& $\ket 1 = 2\sqrt{C_{\varphi^4}}\ket{\phi_1} + \sqrt{-(C^{(1)}_{H^2\varphi^2} + C^{(2)}_{H^2\varphi^2}) }\ket{\varphi}$,\\
	$\qquad \geq - \left(C^{(1)}_{H^2\varphi^2} + C^{(2)}_{H^2\varphi^2}\right)$
	& $\ket 2 =\ket 1$
	\\\hline
	\multirow{2}{*}{$2\sqrt{(C^{(1)}_{H^4}  + C^{(2)}_{H^4} + C^{(3)}_{H^4})C_{\varphi^4}} \geq C^{(2)}_{H^2\varphi^2}$}
	& $\ket 1  = 2\sqrt{C_{\varphi^4}}\ket{\phi_1} + \sqrt{C^{(2)}_{H^2\varphi^2} }\ket{\varphi}$,
	\\
	& $\ket 2 = -2\sqrt{C_{\varphi^4}}\ket{\phi_1} + \sqrt{C^{(2)}_{H^2\varphi^2}}\ket{\varphi}$\\
	\hline
	\end{tabular}
	\caption{Positivity bounds on the left column and the corresponding forward elastic scattering channels on the right column.}
	\label{tab:bounds_channels}
\end{table}

Denoting $A\equiv \sqrt{(C^{(1)}_{H^4}  + C^{(2)}_{H^4} + C^{(3)}_{H^4})C_{\varphi^4}}$, we can rewrite the positivity condition in Eq.~\eqref{eq:boundf} as $-C^{(1)}_{H^2\varphi^2}-2A\leq C^{(2)}_{H^2\varphi^2}\leq 2A$.
As a result, for $C^{(2)}_{H^2\varphi^2}=+1$ and $C^{(1)}_{H^2\varphi^2}\geq 0$, it is interesting to find that the positivity condition in Eq.~\eqref{eq:boundf} leads to $A\geq \frac{1}{2}$, which sets the lower bound on the product of the dimension-8 derivative self-interactions for Higgs and scalar dark matter. On the other hand, for $C^{(2)}_{H^2\varphi^2}=-1$, we get the positivity condition in Eq.~\eqref{eq:boundf} as $C^{(1)}_{H^2\varphi^2}\geq 1-2A$. In this case, the dimension-8 self-interactions for Higgs and scalar dark matter can be small, being compatible with the positivity bounds. Therefore, we focus on the case with $C^{(2)}_{H^2\varphi^2}<0$ in the later discussion.

In the case with the massive spin-2 particle and the radion in Section 2,  from Eqs.~(\ref{t1}), (\ref{t3}) and (\ref{t4}), we can check the positivity bounds, as in the following,
\begin{align}
&C^{(1)}_{H^4}  + C^{(2)}_{H^4}=\frac{2c^2_H\Lambda^4}{m^2_G M^2} \geq 0, \label{ll1} \\ 
&C^{(1)}_{H^4}  + C^{(2)}_{H^4} + C^{(3)}_{H^4} =\frac{4c^2_H\Lambda^4}{3m^2_G M^2}+ \frac{(c^r_H)^2\Lambda^4}{3m^2_r M^2} \geq 0, \label{ll2} \\ 
&C^{(2)}_{H^4} =\frac{c^2_H\Lambda^4}{m^2_G M^2} \geq 0,  \label{ll3} \\
&C^{(1)}_{H^2\varphi^2} =\frac{2c_H c_\varphi\Lambda^4}{m^2_GM^2}\geq 0,\quad {\rm for}\quad c_Hc_\varphi\geq 0, \label{l0} \\
&C_{\varphi^4}=\frac{c^2_\varphi\Lambda^4}{3m^2_G M^2}+\frac{(c^r_\varphi)^2\Lambda^4}{12m^2_r M^2} \geq 0, \label{ll4} \\
&2\sqrt{(C^{(1)}_{H^4}  + C^{(2)}_{H^4} + C^{(3)}_{H^4})C_{\varphi^4}}
\geq -\Big(C^{(1)}_{H^2\varphi^2} + C^{(2)}_{H^2\varphi^2}\Big)=-\frac{4c_Hc_\varphi\Lambda^4}{3m^2_GM^2}-\frac{c^r_Hc^r_\varphi\Lambda^4}{3m^2_rM^2}, \label{l1} \\
&2\sqrt{(C^{(1)}_{H^4}  + C^{(2)}_{H^4} + C^{(3)}_{H^4})C_{\varphi^4}}
\geq C^{(2)}_{H^2\varphi^2}=-\frac{2c_Hc_\varphi\Lambda^4}{3m^2_G M^2}+\frac{c^r_Hc^r_\varphi\Lambda^4}{3m^2_rM^2}. \label{l2}
\end{align}
First, we note that Eqs.~(\ref{ll1}), (\ref{ll2}), (\ref{ll3}) and (\ref{ll4}) are trivially satisfied. Then, as far as $c_Hc_\varphi\geq 0$, we obtain $C^{(1)}_{H^2\varphi^2} \geq 0$ in Eq.~(\ref{l0}), and the last two nontrivial conditions in Eqs.~(\ref{l1}) and (\ref{l2}) are also satisfied automatically. 

Before concluding this section, we also remark the effects of loop corrections on the positivity bounds. In particular, the dimension-6 operators present in the effective theory can also contribute to the Wilson coefficients of the dimension-8 operators at one-loop \cite{Li:2022aby}. Focusing on the Higgs-portal dimension-8 operators, we find that the one-loop corrections give rise to the shifts in the renormalized Wilson coefficients in dimensional regularization, as follows,
\bea
 {\hat C}^{(1)}_{H^2\varphi^2} &=&  C^{(1)}_{H^2\varphi^2}+\frac{1}{648 \pi^2 \Lambda^4}\,\Big(13 ({\tilde d}^2_3+{\tilde d}^2_4)+20  {\tilde d}_3 {\tilde d}_4 \Big) \nonumber \\
 &&+\frac{1}{108\pi^2 \Lambda^4} ({\tilde d}_3+{\tilde d}_4)^2\ln \frac{\mu^2}{|s|}, \\
 {\hat C}^{(2)}_{H^2\varphi^2} &=&  C^{(2)}_{H^2\varphi^2}-\frac{5}{1296\pi^2 \Lambda^4}\, ({\tilde d}_3 + {\tilde d}_4)^2   \nonumber \\
 &&- \frac{1}{432\pi^2 \Lambda^4}\,({\tilde d}_3 + {\tilde d}_4)^2\ln \frac{\mu^2}{|s|}
\eea
where ${\tilde d}_3\equiv d_3 m^2_\varphi$ and ${\tilde d}_4\equiv d_4 m^2_H$ with $d_3, d_4$ given in Eq.~(\ref{lag1}), $\mu$ is the renormalization scale. Here, we took the four-momenta for the four-point vertex  to $\varphi(k)-\varphi(k')-H(p)-H^\dagger(p')$ where $k,k',p$ are incoming toward the vertex and $p'$ is outgoing from the vertex. We also chose the limit of $k=k'$ and $s=(p+k)^2$ is assumed to be spacelike in the above results for simplicity.
Then, the Wilson coefficients for both Higgs-portal dimension-8 operators are corrected due to the Higgs-portal dimension-6 operators, so the positivity bounds for the Higgs-portal dimension-8 operators in Eqs.~(\ref{eq:portal1}) and (\ref{eq:boundf}) are modified by those with $ C^{(1)}_{H^2\varphi^2},  C^{(2)}_{H^2\varphi^2} $ being replaced by ${\hat C}^{(1)}_{H^2\varphi^2},  {\hat C}^{(2)}_{H^2\varphi^2} $, respectively. Thus, for the cutoff scale $\Lambda$ parametrically larger than the dark matter mass considered in the later discussion, the loop corrections renormalized at $\mu=\Lambda$ give rise to small modifications in the positivity bounds, so it is sufficient to consider the positivity bounds on the dimension-8 operators at tree level as discussed in this section. 

We also note that the dimension-6 operators with Higgs fields only can also modify the dimension-8 operators in the SM, but we only have to replace $C^{(1)}_{H^4}, C^{(2)}_{H^4}, C^{(3)}_{H^4} $ by the shifted ones \cite{Li:2022aby}.

\section{Phenomenological constraints}\label{sec:dm_direct-indirect}


In this section, we consider various constraints on the effective Higgs-portal couplings from dark matter relic density, direct and indirect detection as well as the LHC bounds. We show the interplay of direct detection, relic density and positivity bounds in constraining the dimension-8 derivative Higgs-portal couplings and the other Higgs-portal couplings.

\subsection{Dark matter relic density}\label{sec:dm_relic}

In order to determine the relic density by a freeze-out mechanism, we need to consider the annihilation channels for scalar dark matter with the effective Higgs-portal interactions. 

\begin{figure}[!t]
\begin{center}
\includegraphics[width=0.65\textwidth,clip]{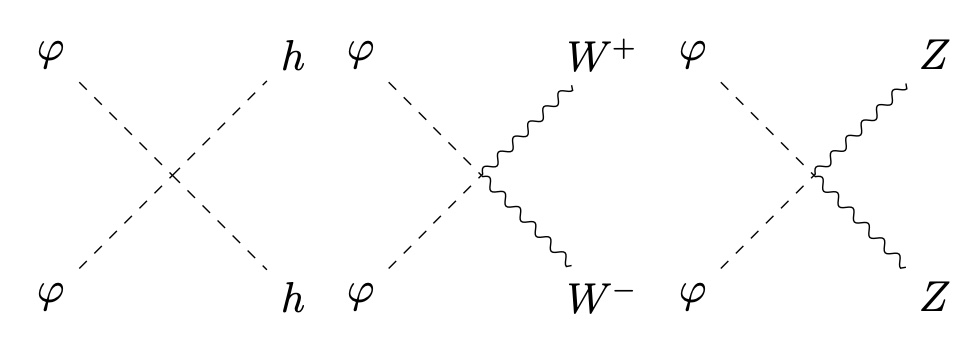}
\end{center}
\caption{Feynman diagrams for dark matter annihilations due to dimension-6 and dimension-8 operators. }
\label{fig:diagram_RelicD}
\end{figure}

For non-derivative Higgs-portal couplings with $c'_3\neq c_3$ and $d'_4\neq d_4$ in Eq.~\eqref{lag1} , there are tree-level Higgs exchanges for dark matter annihilation as in usual Higgs-portal scenarios, such as $\varphi\varphi\to f{\bar f}, VV, hh$ with $f$ being the SM fermions and $V=W, Z$.
For $c'_3=c_3$ and $d'_4=d_4$, the dark matter annihilations, $\varphi\varphi\to f{\bar f}$, are absent at tree level, whereas derivative Higgs-portal couplings for dark matter annihilation coming from Higgs-portal dim-6 and dim-8 interactions, contribute to the other processes for dark matter annihilation, as shown in Fig.~\ref{fig:diagram_RelicD}.

The number density for the dark matter $n_\varphi$ is governed by the following Boltzmann equation,
\begin{align}
\dot{n}_{\varphi} + 3Hn_{\varphi} &= - \langle\sigma v_\mathrm{rel}\rangle_\mathrm{eff}
\left(n^2_\varphi - (n^\mathrm{eq}_\varphi)^2\right),
\label{eq:boltzmann}
\end{align}
where
\bea
\langle\sigma v_\mathrm{rel}\rangle_\mathrm{eff} &= 2\langle\sigma v_\mathrm{rel}\rangle_{\varphi \varphi \to h h}
+ 2\langle\sigma v_\mathrm{rel}\rangle_{\varphi \varphi \to W^+ W^-}
+ 2\langle\sigma v_\mathrm{rel}\rangle_{\varphi \varphi \to ZZ } +2\langle\sigma v_\mathrm{rel}\rangle_{\varphi \varphi \to f{\bar f}},
\eea
$n^\mathrm{eq}_\varphi$ is the number density in thermal equilibrium, and the annihilation cross sections are given by
\begin{align}
\langle\sigma v_\mathrm{rel}\rangle_{\varphi \varphi \to i j} &= \frac{|\mathcal{M}_{\varphi \varphi \to ij}|^2}{32\pi m^2_\varphi}\sqrt{1-\frac{m^2_i}{m^2_\varphi}}.
\end{align}
Here, $(i,j)$ denotes $(h,h)$, $(W^+, W^-)$, $(Z, Z)$, $(f,{\bar f})$, respectively, $m_i$ is the mass of $i$ field, and the $s$-wave contribution is dominant for the annihilation cross sections in the limit $v_\mathrm{rel} \to 0$.
The squared matrix element, $|\mathcal{M}_{\varphi \varphi \to ij}|^2$, includes the symmetric factor 
for identical particles in initial states (i.e. a dark matter particle pair $\varphi \varphi$) and final  states (i.e. $h h$ and $Z Z$).
Then, we get the squared scattering matrix elements for electroweak final states as
\bea
|\mathcal{M}_{\varphi \varphi \to h h}|^2 &= &\frac{m^4_{\varphi}}{9\Lambda^8 (m^2_h - 4m^2_{\varphi})^2}
\bigg[\Big(2 c_3 - d_3 + d_4 + 3 C^{(2)}_{H^2\varphi^2} \Big) m^4_h  \nonumber \\
&& + \Big(4 c_3 - 12 c'_3 + 6 d_3 + 2 d_4 - 6 d'_4 - 3 C^{(1)}_{H^2\varphi^2}  - 18 C^{(2)}_{H^2\varphi^2} \Big) m^2_h m^2_{\varphi}  \nonumber \\
&&+ 4 \Big( - 2 d_3 + 3  C^{(1)}_{H^2\varphi^2} + 6  C^{(2)}_{H^2\varphi^2}\Big) m^4_{\varphi}\bigg]^2,
\label{eq:dmdm_hh}
\eea
\bea
|\mathcal{M}_{\varphi \varphi \to W^+ W^-}|^2 &=& \frac{2\pi^2\alpha^2m^4_{\varphi}v^4}{9\Lambda^8 m^4_W s^4_W}
\bigg[
9 (C^{(1)}_{H^2\varphi^2})^2 (m^2_{\varphi} - m^2_W)^2  - \frac{6 C^{(1)}_{H^2\varphi^2}}{m^2_h - 4 m^2_{\varphi}}  (m^2_{\varphi}  - m^2_W) (2m^2_{\varphi}  - m^2_W)\nonumber   \\
&&\times \left\{\Big(2 (c_3 - c'_3) + d_3 + d_4 - d'_4 - 3 C^{(2)}_{H^2\varphi^2}\Big) m^2_h+ 4 \Big(- d_3 + 3 C^{(2)}_{H^2\varphi^2}\Big) m^2_{\varphi}  \right\} \nonumber \\
&&+ \frac{1}{(m^2_h - 4 m^2_{\varphi})^2}(4 m^4_{\varphi}  - 4 m^2_{\varphi}  m^2_W + 3 m^4_W)\nonumber  \\
&& \times \left\{\Big(2 (c_3 - c'_3) + d_3 + d_4 - d'_4 - 3 C^{(2)}_{H^2\varphi^2}\Big) m^2_h+ 4 \Big(- d_3 + 3 C^{(2)}_{H^2\varphi^2}\Big) m^2_{\varphi} \right\}^2 \bigg], \nonumber \\
\label{eq:dmdm_ww}
\eea
\bea
|\mathcal{M}_{\varphi \varphi \to Z Z}|^2=\frac{1}{2} |\mathcal{M}_{\varphi \varphi \to W^+W^-}|^2 (m_W\rightarrow m_Z, s_W\rightarrow s_W c_W)
\label{eq:dmdm_zz}
\eea
where $c_W = \cos{\theta_W}$, $s_W = \sin{\theta_W}$, $\theta_W$ is the Weinberg angle, $\alpha$ is the fine structure constant, and $v$ is the vacuum expectation value of the SM Higgs field. Moreover,  the squared scattering matrix elements for $\varphi \varphi \to f{\bar f}$, with $ f{\bar f} = t{\bar t}, b{\bar b}$, are given by
\begin{align}
\label{eq:dmdm_ff}
|\mathcal{M}_{\varphi \varphi \to f {\bar f}}|^2 &= \frac{4 m^2_f m^4_h m^4_{\varphi} (m^2_{\varphi} - m^2_f)}
{3\Lambda^8 (m^2_h - 4 m^2_{\varphi})^2}\,\cdot  \Big(2(c_3 - c'_3) + d_4 - d'_4\Big)^2,
\end{align}
where $m_f = m_t, m_b$  are top and bottom quark masses. 

Consequently, all the annihilation channels in the effective theory approach, if allowed, are dominated by the $s$-wave in the generic parameter space. This is in contrast with the case with a massive graviton mediator where $\varphi\varphi\to hh, f{\bar f}$ become $d$-wave suppressed and $\varphi\varphi\to VV$ become also $d$-wave suppressed for the universal couplings of the massive graviton to electroweak bosons in the SM including the gauge kinetic terms \cite{Lee:2013bua}. 

For the universal couplings for dimension-4 and dimension-6 operators, $c_3=c'_3=d_3=d_4=d'_4=-3C^{(1)}_{H^2\varphi^2}/2-6 C^{(2)}_{H^2\varphi^2}$, which is the case with graviton and radion from Eq.~(\ref{t2}), we can simplify the squared matrix elements for  $\varphi\varphi\to hh, WW$, as follows,
\bea
|\mathcal{M}_{\varphi \varphi \to h h}|^2&=& \frac{m^4_\varphi}{\Lambda^8}\, (m^2_h+2m^2_\varphi)^2 \Big( C^{(1)}_{H^2\varphi^2}+3C^{(2)}_{H^2\varphi^2}\Big)^2, \label{hhsimp} \\
|\mathcal{M}_{\varphi \varphi \to W^+ W^-}|^2&=& \frac{2\pi^2\alpha^2m^4_{\varphi}v^4}{9\Lambda^8 m^4_W s^4_W}\bigg[36 (C^{(1)}_{H^2\varphi^2} + 3 C^{(2)}_{H^2\varphi^2})^2 m_\varphi^4 \nonumber \\
&&- 54 (C^{(1)}_{H^2\varphi^2} + 2 C^{(2)}_{H^2\varphi^2}) (C^{(1)}_{H^2\varphi^2} + 3 C^{(2)}_{H^2\varphi^2}) m_\varphi^2 m_W^2 \nonumber \\
&&+ \frac{9}{4} \Big(11 (C^{(1)}_{H^2\varphi^2})^2 + 60 C^{(1)}_{H^2\varphi^2} C^{(2)}_{H^2\varphi^2} + 108 (C^{(2)}_{H^2\varphi^2})^2\Big) m_W^4\bigg],
\eea
and the squared matrix elements for  $\varphi\varphi\to f{\bar f}$ vanish. With the graviton only, we can take a further condition, $C^{(2)}_{H^2\varphi^2}=-C^{(1)}_{H^2\varphi^2}/3$, from Eq.~(\ref{G1}). In this case, the squared matrix elements for $\varphi\varphi\to hh$ also vanish at $s$-wave \cite{Lee:2013bua}, but $\varphi\varphi\to VV$ is $s$-wave dominant, because
\bea
|\mathcal{M}_{\varphi \varphi \to W^+ W^-}|^2= \frac{2\pi^2\alpha^2m^4_{\varphi}v^4}{9\Lambda^8 s^4_W}\,  (C^{(1)}_{H^2\varphi^2})^2. \label{vvsimp}
\eea
Therefore, in the effective theory stemming from the massive graviton, there is no strong bound from indirect detection experiments for either $\varphi\varphi\to hh$ or $\varphi\varphi\to f{\bar f}$  \cite{Lee:2013bua,Lee:2014caa}.

By solving the Boltzmann equation in Eq.~\eqref{eq:boltzmann}, we can obtain the abundance for dark matter in terms of $Y_\mathrm{DM} = n_{\varphi}/s$ at present, as follows,
\begin{align}
\Omega_\mathrm{DM} h^2 = 0.2744 \left(\frac{Y_\mathrm{DM}}{10^{-11}}\right)\left(\frac{m_{\varphi}}{100~\mathrm{GeV}}\right).
\end{align}
The present abundance for dark matter is approximated \cite{Cline:2013gha, Cline:2018fuq, Scherrer:1985zt} to
\begin{align}
Y_\mathrm{DM} \simeq \sqrt{\frac{45g_*}{\pi g^2_{*s}}}\frac{x_f}{m_{\varphi}M_\mathrm{pl}\langle\sigma v_\mathrm{rel}\rangle_{\mathrm{eff}}},\label{eq:YDM_analytic}
\end{align}
where 
\bea
x_f \simeq \ln{y_f} - \frac{1}{2}\ln\ln y_f, \quad y_f =\frac{1}{2\pi^3}\sqrt{\frac{45}{8g_*}} m_{\varphi} M_\mathrm{pl}\langle\sigma v_\mathrm{rel}\rangle_{\mathrm{eff}},
\eea
where $M_\mathrm{pl} = 1.22\times10^{19}~$GeV is the Planck mass, and $g_*, g_{*s}$ are the effective numbers of relativistic degrees of freedom in radiation and entropy, respectively.

\begin{figure}[!t]
\begin{center}
 \includegraphics[width=0.40\textwidth,clip]{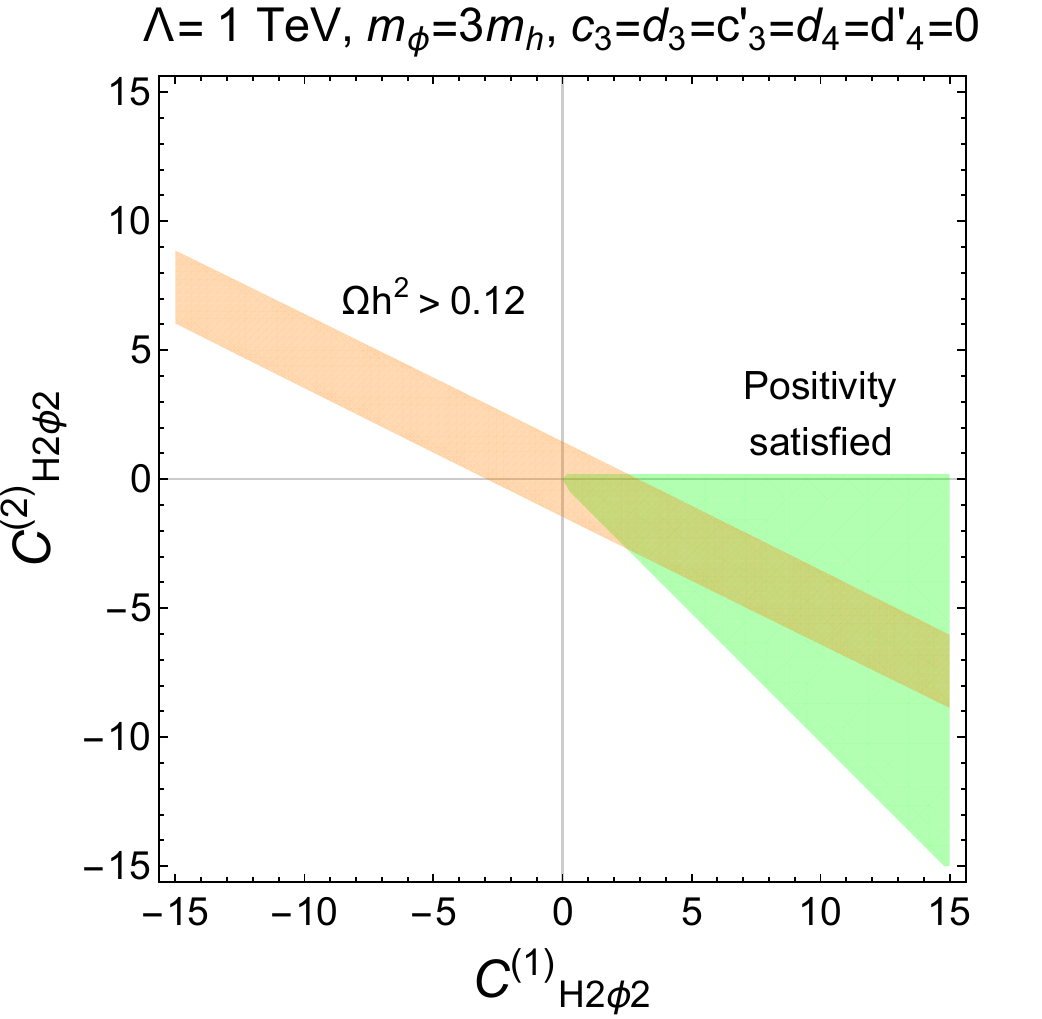} \,\,  \includegraphics[width=0.40\textwidth,clip]{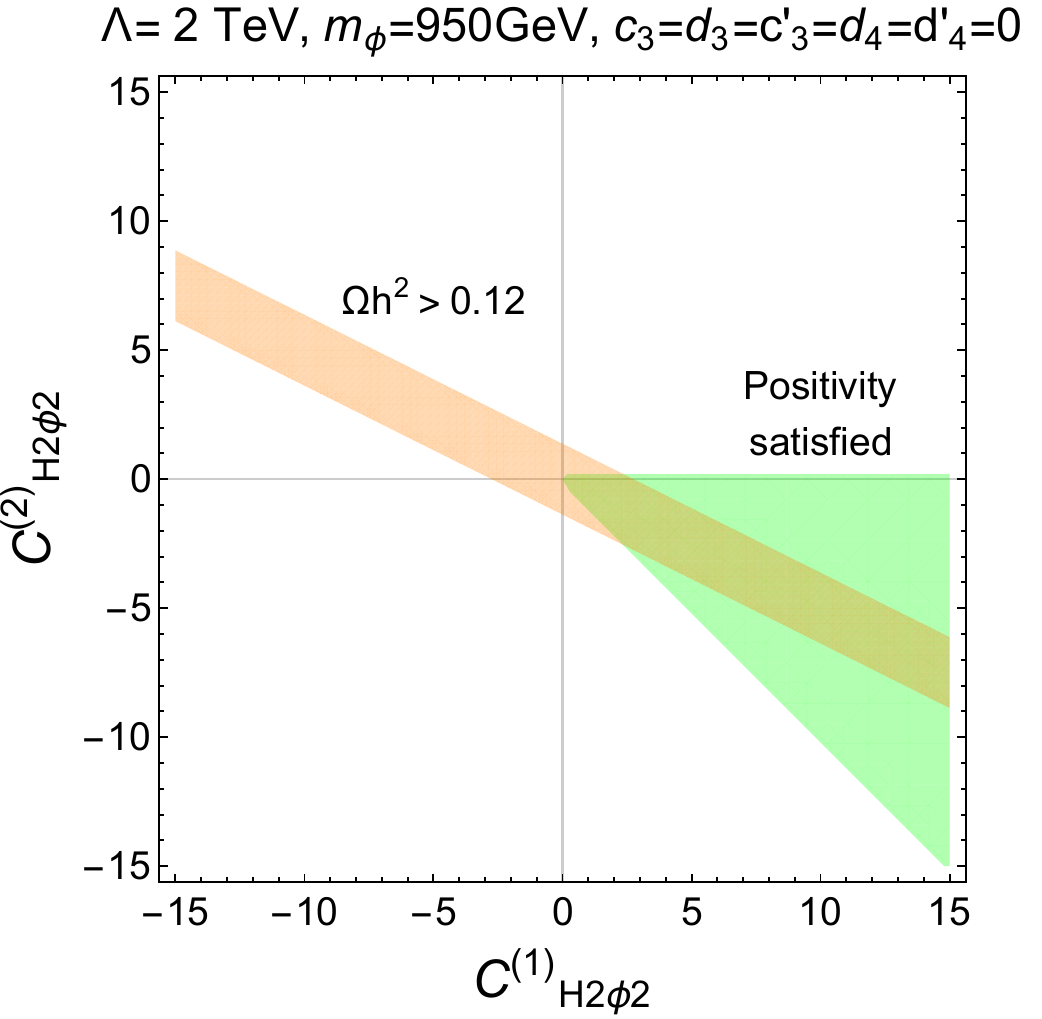} \vspace{0.5cm} \\
  \includegraphics[width=0.40\textwidth,clip]{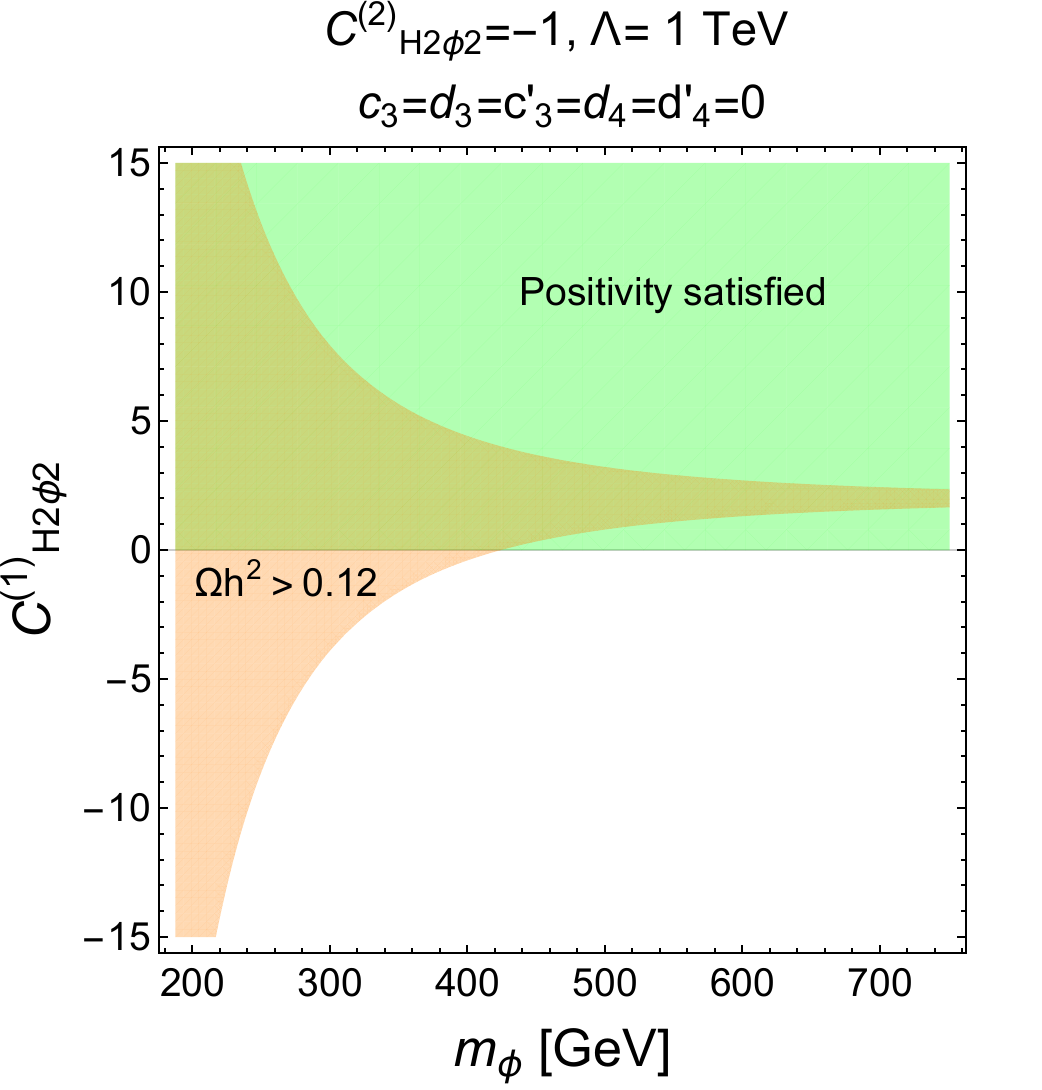} \,\,  \includegraphics[width=0.40\textwidth,clip]{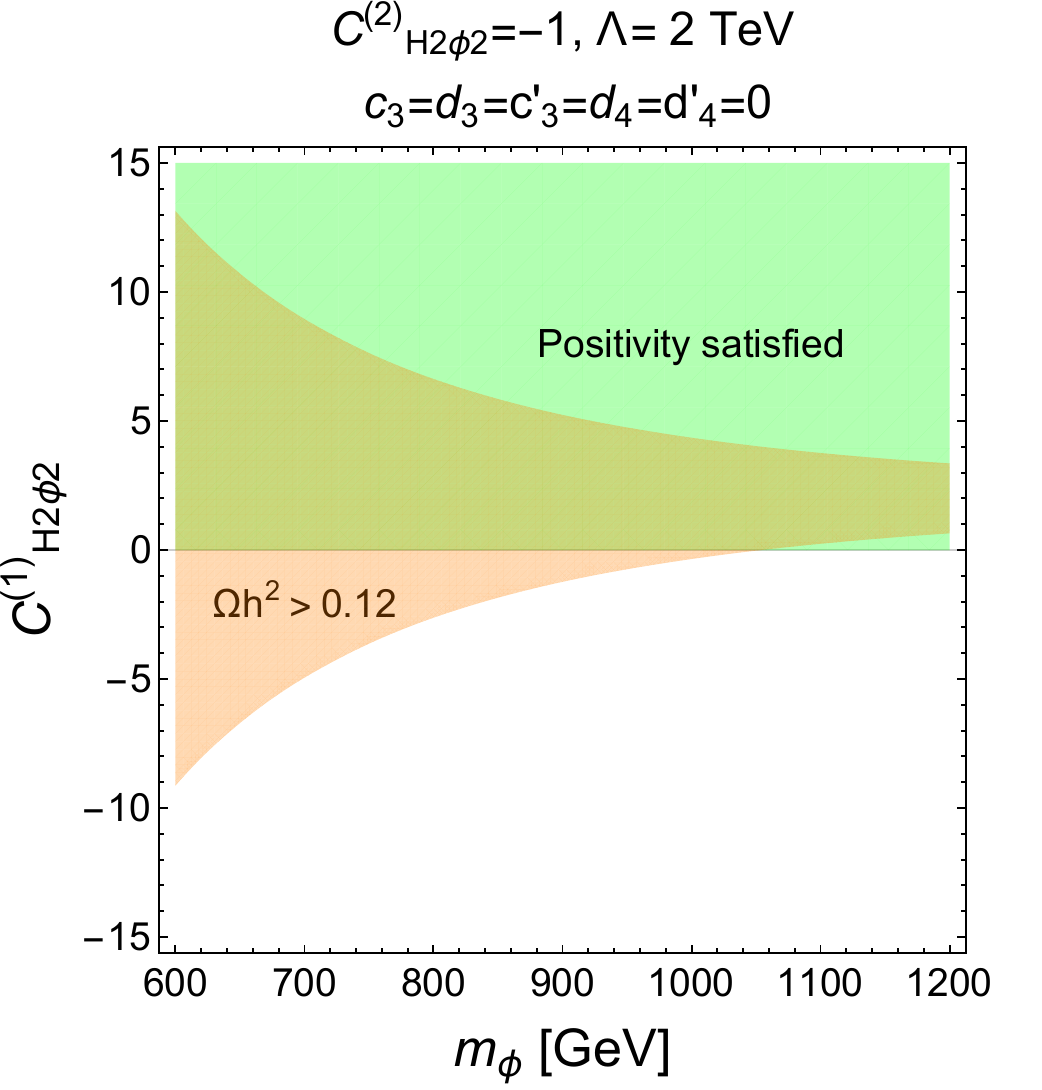}
\end{center}
\caption{Upper: Parameter space for $C^{(1)}_{H^2\varphi^2}$ vs $C^{(2)}_{H^2\varphi^2}$, satisfying positivity and relic density.
We took $A\equiv \sqrt{(C^{(1)}_{H^4}  + C^{(2)}_{H^4} + C^{(3)}_{H^4})C_{\varphi^4}}=0.1$ and $m_{\varphi} = 3m_h (950\,{\rm GeV}) $ for the left(right) figures.
Lower: Parameter space for $m_{\varphi}$ vs $C^{(1)}_{H^2\varphi^2}$, satisfying positivity and relic density, for $C^{(2)}_{H^2\varphi^2}=-1$. We took $\Lambda = 1(2)$ TeV for the left(right) figures in each panel. The relic density for dark matter is overproduced in orange regions, namely, $\Omega h^2>0.12$, and it saturates the observed value along the boundary of the orange region. The positivity bounds in Eq.~\eqref{eq:portal1} are satisfied in green regions. In all the plots, we set $c_3=d_3=c'_3=d_4=d'_4=0$. }
\label{fig:RelicD}
\end{figure}

\begin{figure}[!t]
\begin{center}
 \includegraphics[width=0.40\textwidth,clip]{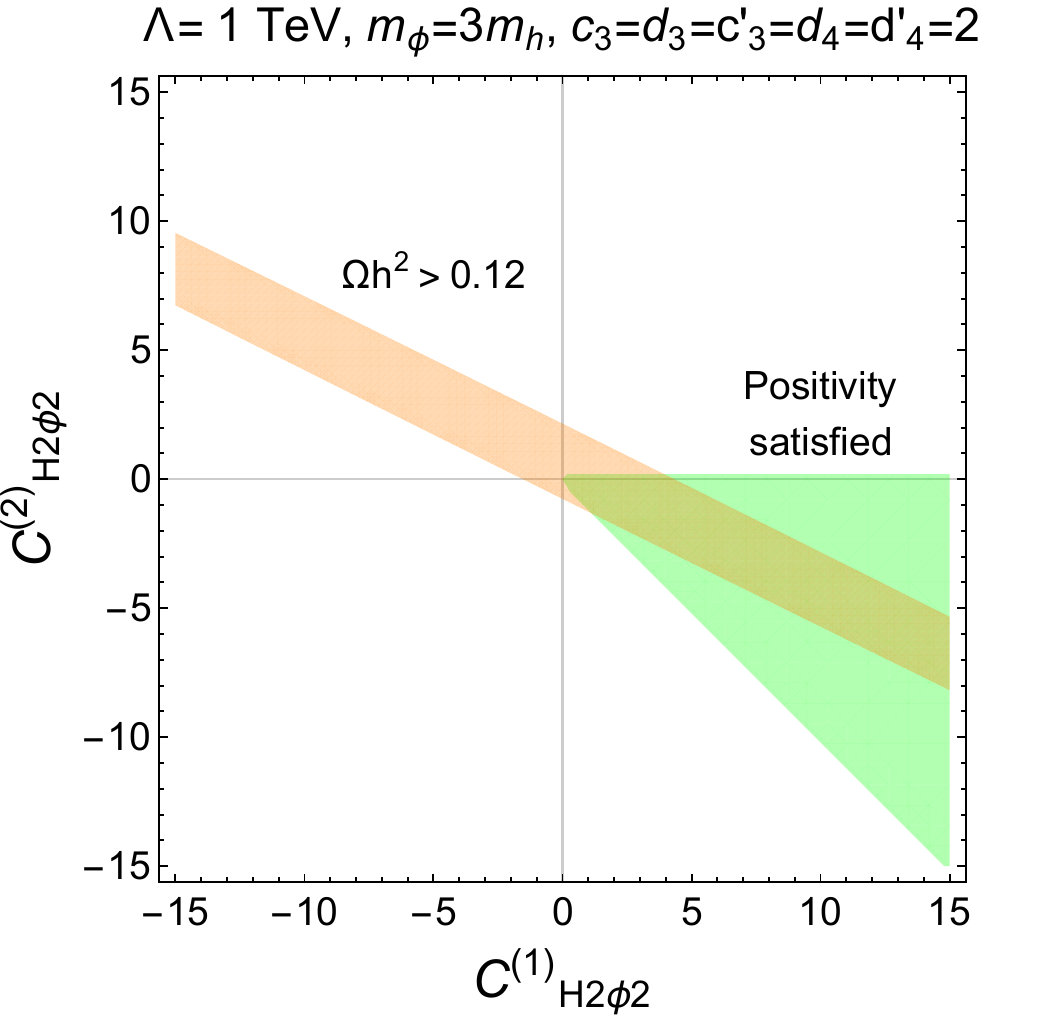} \,\,  \includegraphics[width=0.40\textwidth,clip]{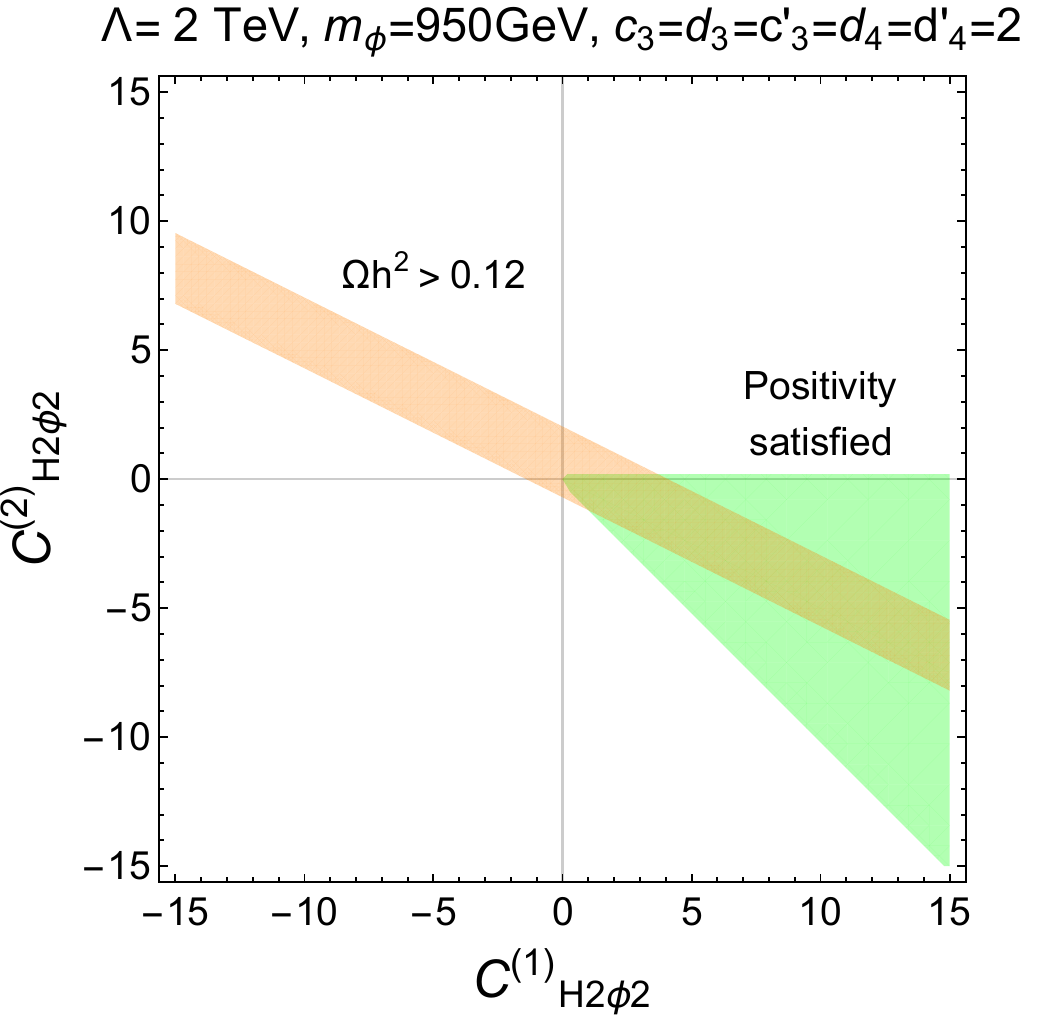} \vspace{0.5cm} \\
  \includegraphics[width=0.40\textwidth,clip]{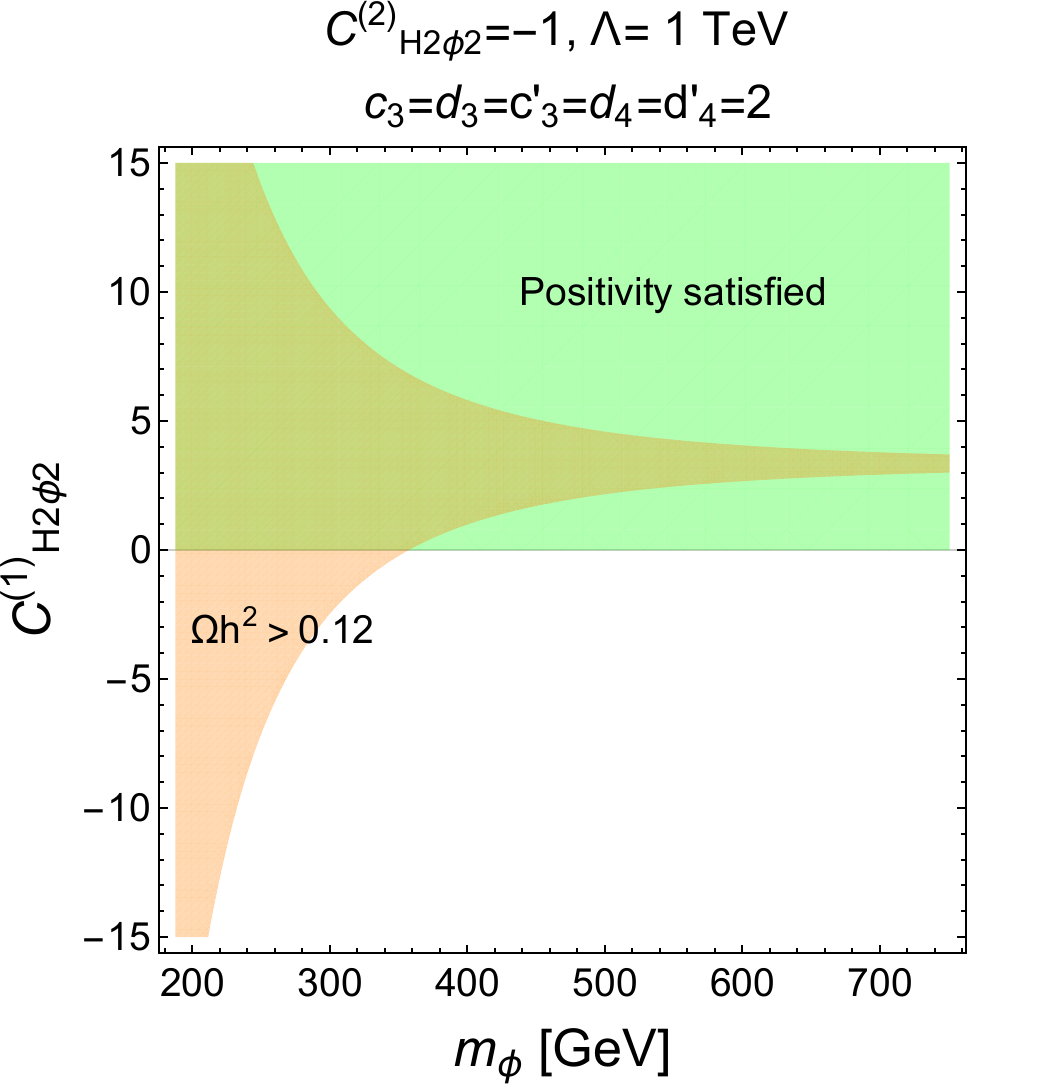} \,\,  \includegraphics[width=0.40\textwidth,clip]{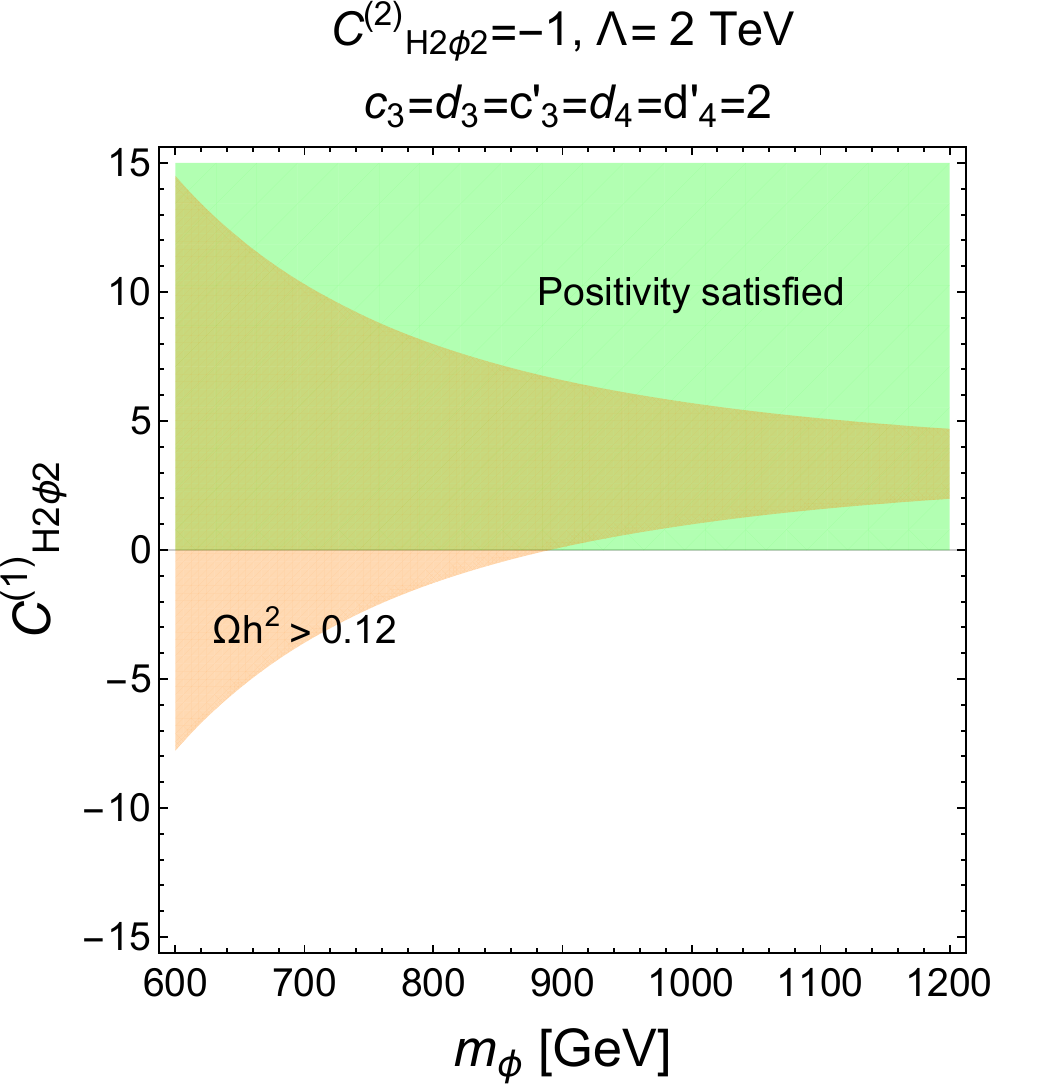}
\end{center}
\caption{The same as in Fig.~\ref{fig:RelicD}, but with $c_3=d_3=c'_3=d_4=d'_4=2$. }
\label{fig:RelicD2}
\end{figure}

\begin{figure}[!t]
\begin{center}
 \includegraphics[width=0.40\textwidth,clip]{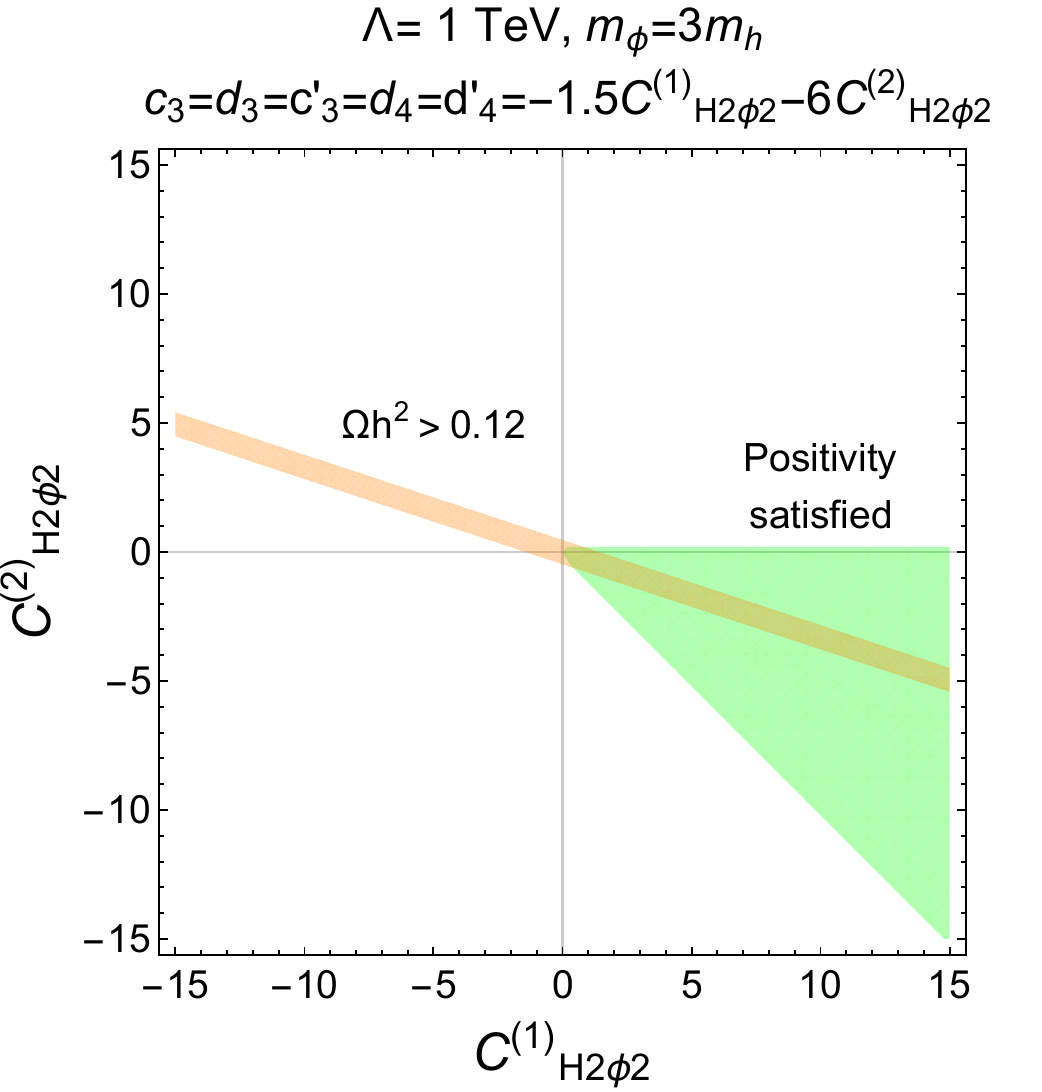} \,\,  \includegraphics[width=0.40\textwidth,clip]{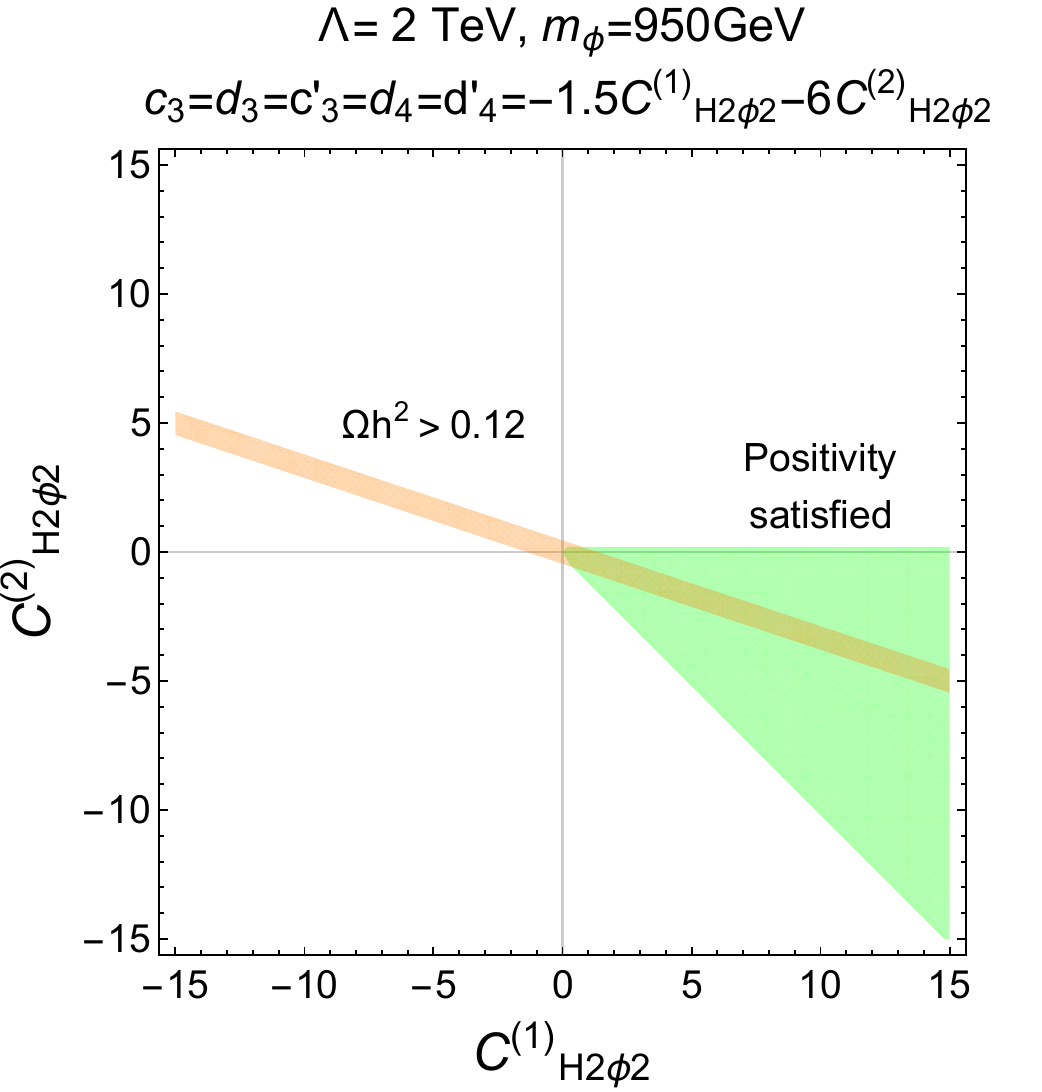} \vspace{0.5cm} \\
  \includegraphics[width=0.40\textwidth,clip]{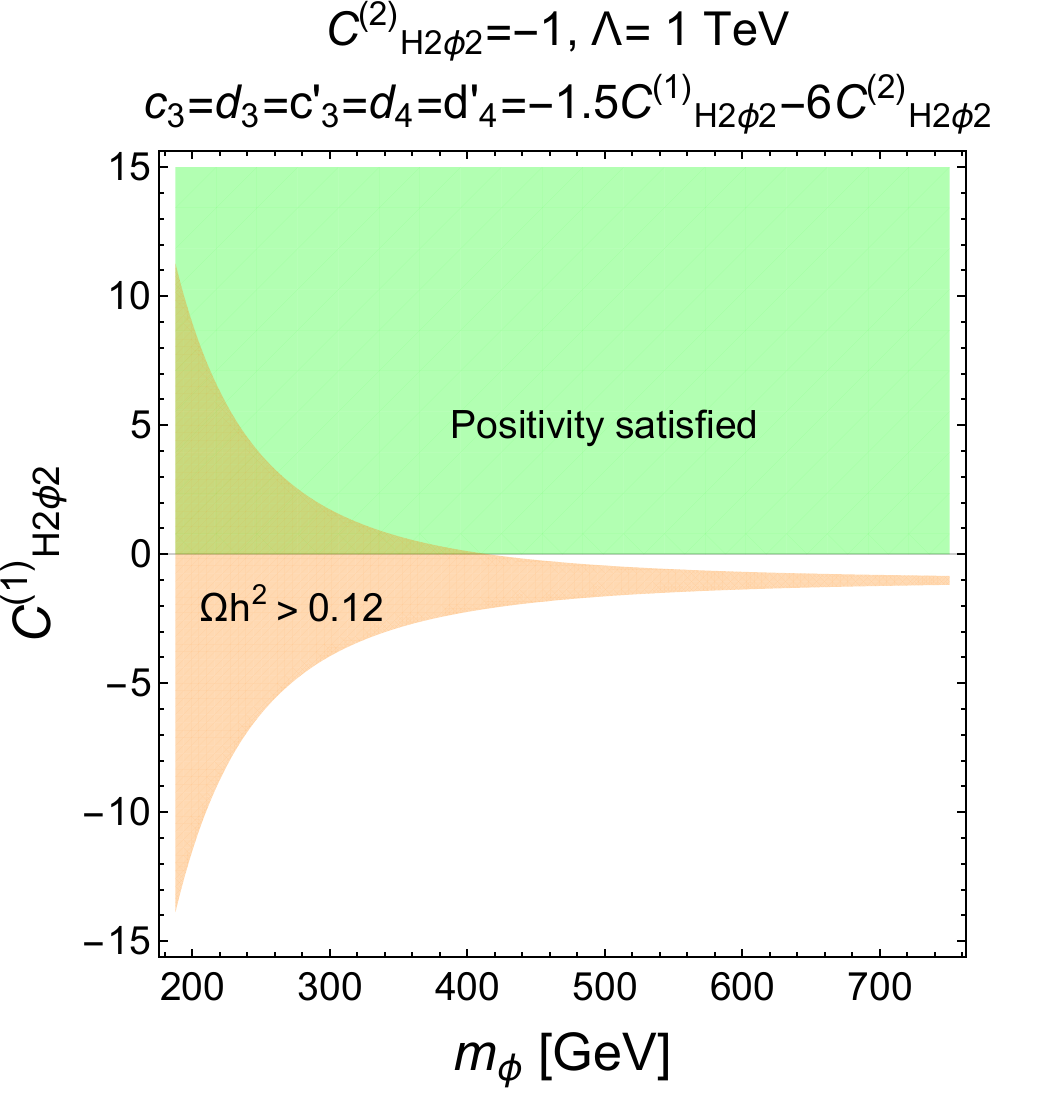} \,\,  \includegraphics[width=0.40\textwidth,clip]{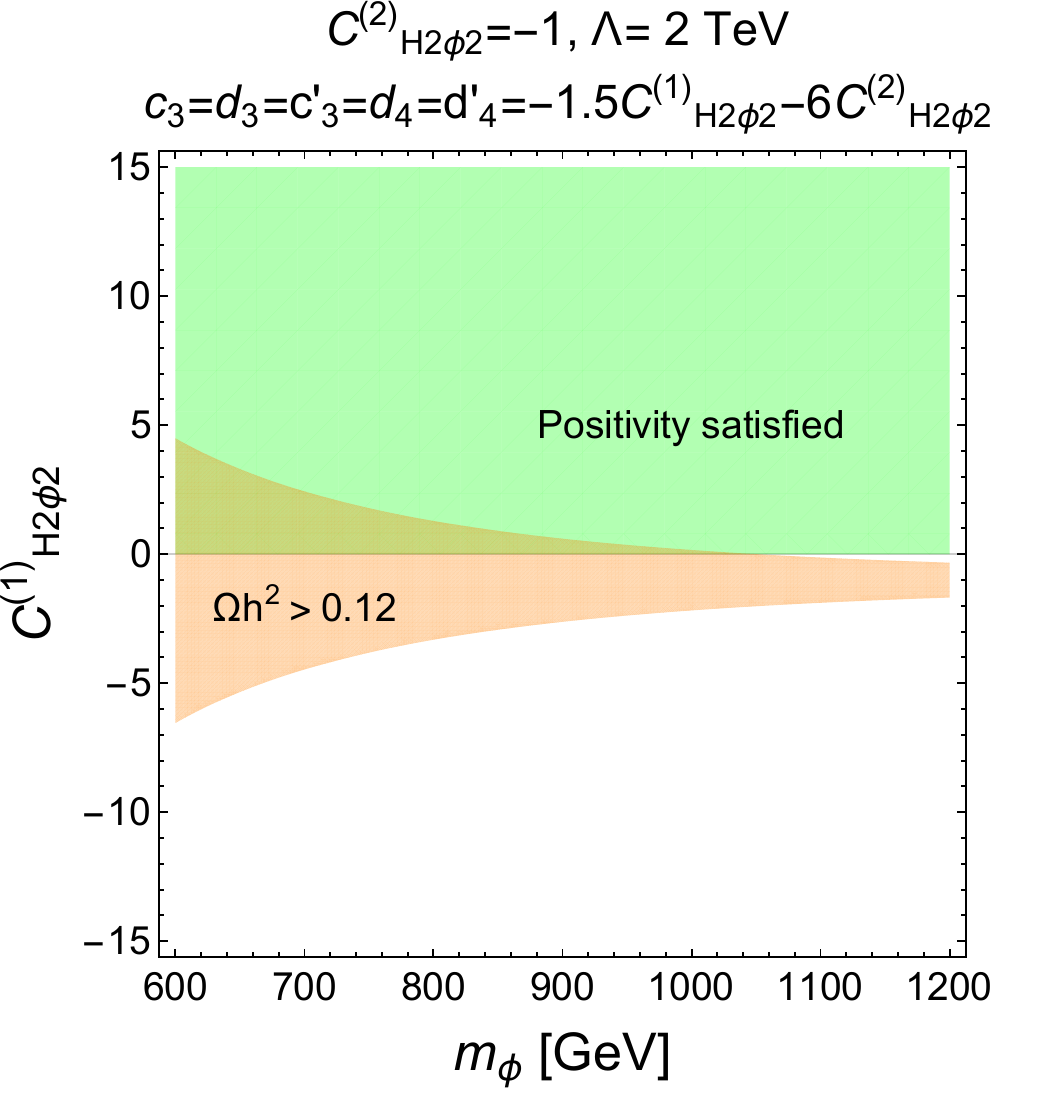}
\end{center}
\caption{The same as in Fig.~\ref{fig:RelicD}, but with $c_3=d_3=c'_3=d_4=d'_4=-1.5 C^{(1)}_{H^2\varphi^2}-6C^{(2)}_{H^2\varphi^2}$ for graviton and radion. }
\label{fig:RelicD3}
\end{figure}

\subsection{Direct detection}

The effective Higgs-portal couplings, $c_3$, $c'_3$, $d_4$, and $d'_4$, give rise to the effective interactions between dark matter and quarks through Higgs boson exchanges in the $t$-channel, as follows,
\begin{align}
{\cal L}_{{\rm eff}, \varphi-q}= - \frac{(2(c_3 - c'_3) - d_4 + d'_4) m_q m^2_\varphi}{6\Lambda^4} \varphi^2 {\bar q}q.
\end{align}
Then, we obtain the cross section for the spin-independent scattering between dark matter and nucleus, as follows,
\bea
\sigma_{\varphi-X} =\frac{\mu^2_N}{\pi m^2_\varphi A^2} \Big(Z f_p +(A-Z)f_n \Big)^2 \label{DD}
\eea
where $Z, A-Z$ are the number of protons and neutrons in the detector nucleus, $\mu_X=m_\varphi m_X/(m_\varphi+m_X)$ is the reduced mass for the DM-nucleus system, and the nucleon form factors are given by
\begin{align}
f_{p,n} =-\frac{(2(c_3 - c'_3) - d_4 + d'_4) m_{p,n} m^2_\varphi }{6\Lambda^4} \bigg(\sum_{q=u,d,s} f^{p,n}_{Tq}+ \frac{2}{9} f^{p,n}_{TG} \bigg),
\end{align}
with $f^{p,n}_{TG}=1-\sum_{q=u,d,s} f^{p,n}_{Tq}$. Here, $f^N_{Tq}$ is the mass fraction of quark $q$ inside the nucleon $N$, defined by $\langle N|m_q {\bar q}q |N\rangle= m_N f^N_{Tq}$, and $f^{N}_{TG}$ is the mass fraction of gluon $G$  the nucleon $N$, due to heavy quarks. The numerical values are $f^p_{T_u}=0.0208\pm 0.0015$ and $f^p_{T_d}=0.0411\pm 0.0028$ for a proton, $f^n_{T_u}=0.0189\pm 0.0014$ and $f^n_{T_d}=0.0451\pm 0.0027$ for a neutron \cite{DDupdate}, and  $f^{p,n}_{T_s}=0.043\pm 0.011$ for both proton and neutron \cite{DDstrange}.

\begin{figure}[!t]
\begin{center}
 \includegraphics[width=0.40\textwidth,clip]{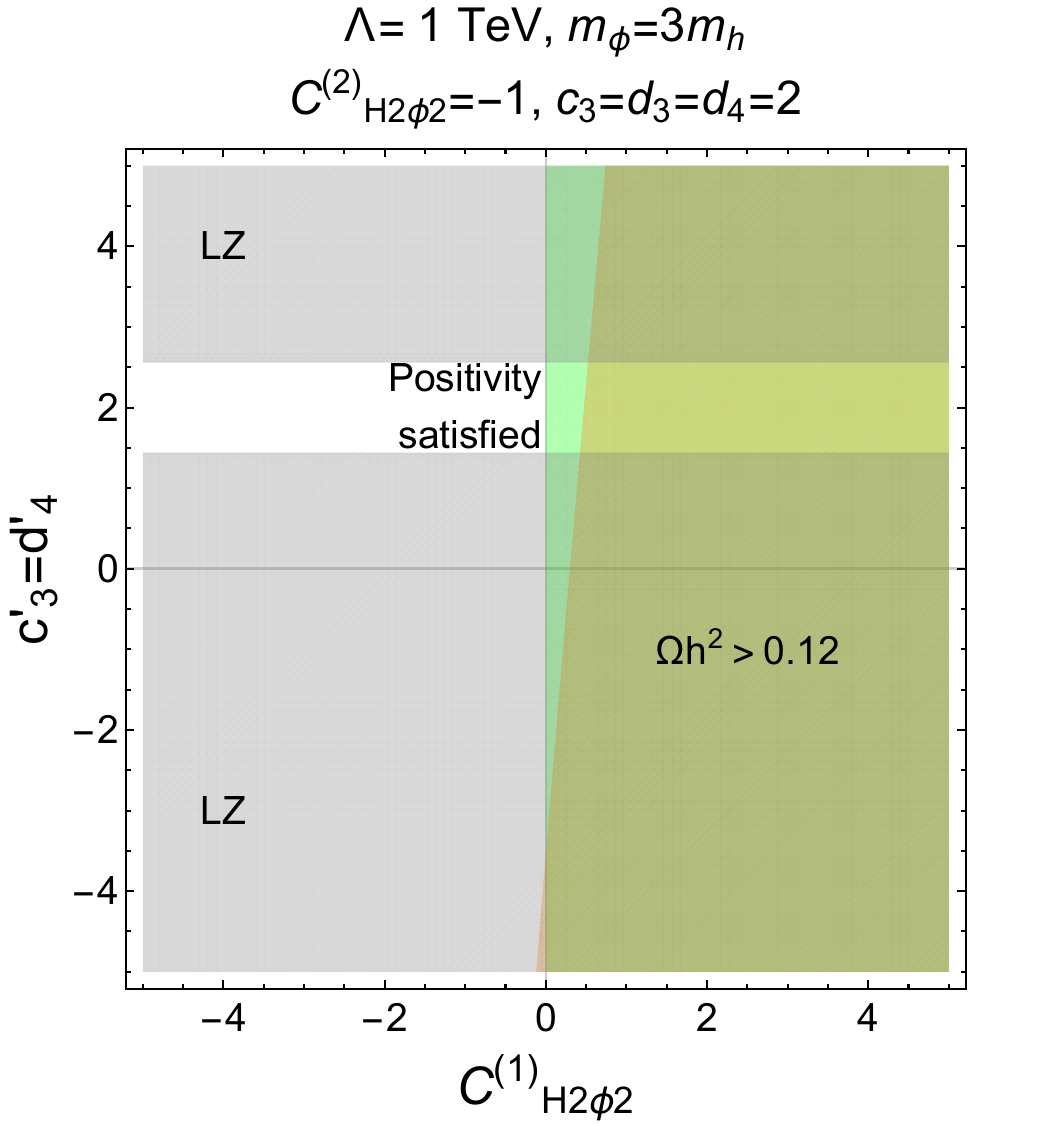} \,\,  \includegraphics[width=0.40\textwidth,clip]{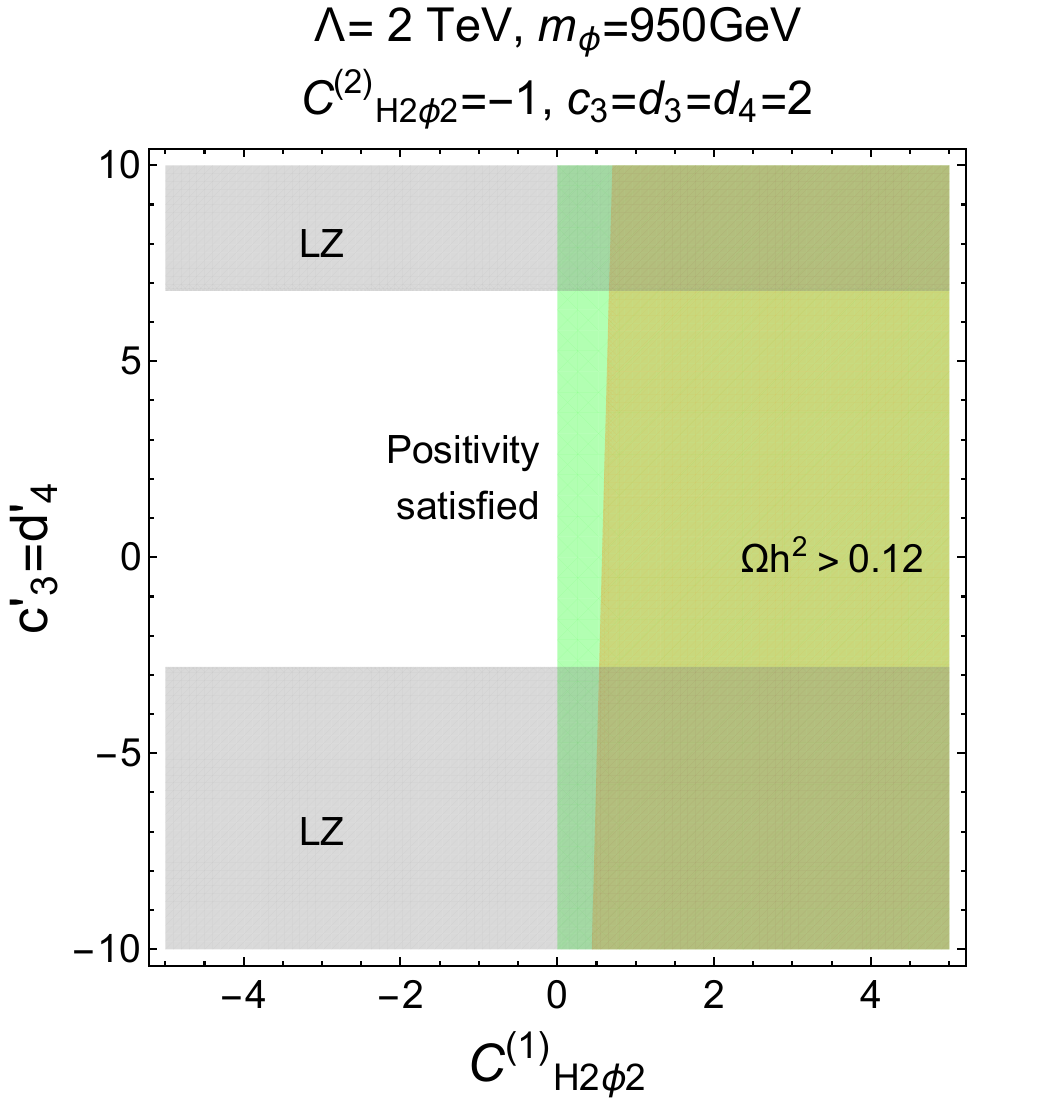} 
\end{center}
\caption{Parameter space for $C^{(1)}_{H^2\varphi^2}$ vs $c'_3=d'_4$. The present relic density for dark matter goes beyond the measured value in orange regions, namely, $\Omega h^2>0.12$. The gray regions are excluded by the LZ experiment \cite{LZ}.  The positivity bounds are satisfied in green regions. We took $\Lambda = 1(2)$ TeV and $m_\varphi=3m_h(950\,{\rm GeV})$ for the left(right) figure, and $c_3=d_3=d_4=2$ and $C^{(2)}_{H^2\varphi^2}=-1$ for both plots.}
\label{fig:DD}
\end{figure}	

Taking the universal Higgs-portal couplings for dimension-4 and dimension-6 operators, $c'_3=c_3$ and $d'_4=d_4$, the momentum transfer independent part of the direct detection cross section in Eq.~(\ref{DD}) vanishes, so there is no strong bound from the direct detection experiments. The effective theory obtained from graviton and radion respects such universal relations for effective couplings, as shown in Eq.~(\ref{t2}), which could be ensured at the cutoff scale. Of course, in the effective theory where we are agnostic about the origin of the effective couplings, the direct detection can constrain only the combination of the effective low-energy couplings to $2(c_3 - c'_3) = d_4 - d'_4$, from Eq.~(\ref{DD}).

In Fig.~\ref{fig:RelicD}, we show the parameter space in $C^{(1)}_{H^2\varphi^2}$ and $C^{(2)}_{H^2\varphi^2}$ in the upper panel and $m_\varphi$ and $C^{(1)}_{H^2\varphi^2}$ with $C^{(2)}_{H^2\varphi^2}=-1$ in the lower panel, satisfying the relic density for dark matter and positivity bounds. We have fixed $\Lambda=1, 2\,{\rm TeV}$ on left and right figures, respectively, and $m_\varphi=3m_h, 950\,{\rm GeV}$ on the left and right figures in the upper panel, respectively. We set dimension-4 and dimension-6 couplings to zero, namely, $c_3=d_3=c'_3=d_4=d'_4=0$, for all the plots in  Fig.~\ref{fig:RelicD}. The relic density with $\Omega h^2<0.12$ is achieved outside the orange regions, so the observed relic density, $\Omega h^2=0.12$, is explained along the boundary of the orange region, and the positivity bounds are satisfied in the green regions. The direct detection bounds for dark matter are satisfied for all the plots. 

For the plots in the upper panel of  Fig.~\ref{fig:RelicD}, we also took the combination of the dimension-8 derivative self-interactions for Higgs and dark matter by $A\equiv \sqrt{(C^{(1)}_{H^4}  + C^{(2)}_{H^4} + C^{(3)}_{H^4})C_{\varphi^4}}=0.1$. However, for given LHC bounds on the Wilson coefficients of the Higgs self-couplings up to order one discussed in Section~\ref{sec:lhc}, we can also allow for a larger value of $A$ because $C_{\varphi^4}$ is unconstrained for WIMP-like dark matter, so the positivity bounds are satisfied even for positive values of $C^{(2)}_{H^2\varphi^2}$ in the upper panel. 

In Figs.~\ref{fig:RelicD2} and \ref{fig:RelicD3}, we present the similar results as in Fig.~\ref{fig:RelicD}, except with $c_3=d_3=c'_3=d_4=d'_4=2$ for the former case and $c_3=d_3=c'_3=d_4=d'_4=-1.5 C^{(1)}_{H^2\varphi^2}-6C^{(2)}_{H^2\varphi^2}$ in the latter case. Dimension-4 and dimension-6 Higgs portal couplings are (un-)correlated to dimension-8 operators in the latter(former) cases, and the latter case can be derived from graviton and radion. Then, in Fig.~\ref{fig:RelicD2}, the relic density condition shifts  the allowed parameter for $C^{(2)}_{H^2\varphi^2}$ towards more positive values, because the dimension-4 and dimension-6 Higgs portal couplings also contribute to the relic density. On the other hand, in Figs.~\ref{fig:RelicD3}, the relic density condition makes the allowed parameter space for $C^{(1)}_{H^2\varphi^2}$ and $C^{(2)}_{H^2\varphi^2}$ narrower, due to the correlation between dimension-8 and lower-dimensional Higgs-portal couplings. Therefore, the relic density condition depends crucially on the presence of the dimension-4 and dimension-6 Higgs portal couplings.
 
 In Fig.~\ref{fig:DD}, we depict the parameter space in $C^{(1)}_{H^2\varphi^2}$ and Higgs-portal couplings with $c'_3=d'_4$, satisfying the relic density and the positivity conditions as well as the direct detection bound from the LUX-ZEPLIN(LZ) experiment \cite{LZ}. We chose $c_3=d_3=d_4=2$ and $C^{(2)}_{H^2\varphi^2}=-1$ for both plots and $\Lambda = 1, 2$ TeV and $m_\varphi=3m_h, 950\,{\rm GeV}$ for the left and right figures, respectively.
For relatively light dark matter on left, the consistent parameter space remains close to the universal couplings, $c_3=d_4=c_3'=d'_4$, outside the gray region excluded by the LZ experiment. On the other hand, for heavy dark matter on right, there is still a lot of parameter space left to be compatible with the direct detection bound. We find that the positivity bounds in green are complementary to the bounds from direct detection in gray and relic density in orange in constraining the Higgs-portal effective interactions.

\subsection{Indirect detection}\label{sec:indirect}

Dark matter can annihilate into $f{\bar f}$, $VV$ with $V=W, Z$ or $hh$ without velocity suppression for the generic parameter space of the effective theory approach. In this case, the effective Higgs-portal couplings can be constrained by indirect detection experiments \cite{indirect,LQ} such as Fermi-LAT dwarf galaxies \cite{fermi-lat}, HESS gamma-rays \cite{hess}, AMS-02 antiprotons \cite{ams}, and Cosmic Microwave Background measurements \cite{SRDM}. 

On the other hand, for universal effective couplings for dimension-4 and dimension-6 operators, as discussed in the previous subsection, $\varphi\varphi\to f{\bar f}$ is absent, so the $s$-wave dominant channels, $\varphi\varphi\to hh, VV$, can be constrained importantly by indirect detection. In particular, the strong bounds from direct detection can be satisfied for $c'_3=c_4$ and $d_4=d'_4$, leading to interesting signatures for indirect detection from heavy bosons in the SM. The dark matter annihilation into heavy gauge bosons has been less constrained by indirect searches \cite{IDWW}, but it is potentially discoverable by the Milky Way Galactic Center from Fermi-LAT \cite{IDGC} and more data with indirect detection. 

If we further impose the correlation between dimension-8 and lower-dimensional Higgs-portal couplings as in Eq.~(\ref{t2}) with Eq.~(\ref{G1}),  the $\varphi\varphi\to hh$ channel is velocity-suppressed, as discussed below Eq.~(\ref{hhsimp}), whereas the $\varphi\varphi\to VV$ channels are still $s$-wave and they could lead to interesting signatures in cosmic ray observations.

\subsection{LHC searches}\label{sec:lhc}

In this subsection, we discuss the current limits on the dimension-8 operators for Higgs only in the third line of Table~\ref{tab:dim8_operators} and discuss the dark matter production at HL-LHC, induced by the dimension-8 Higgs portal operators, i.e, the first line of Table~\ref{tab:dim8_operators}.

First, the dimension-8 derivative self-couplings for the Higgs are constrained most by the same sign $W$ boson pairs at the LHC to $C^{(2)}_{H^4}/\Lambda^4=[-7.7,7.7]\,{\rm TeV}^{-4}$ and $C^{(3)}_{H^4}/\Lambda^4=[-21.6,21.8]\,{\rm TeV}^{-4}$ at 95\% CL \cite{LHCWW} (See also the weaker limits from the $WZ$ boson pairs \cite{LHCWZ}), but there is no limit shown for  $C^{(1)}_{H^4}$, although a similar limit is expected. 
Moreover, the combined $WW, WZ, ZZ$ channels in association with two jets lead to stronger limits, $C^{(2)}_{H^4}/\Lambda^4=[-2.7,2.7]\,{\rm TeV}^{-4}$ and $C^{(3)}_{H^4}/\Lambda^4=[-3.4,3.4]\,{\rm TeV}^{-4}$ at 95\% CL \cite{LHCbosons}.

\begin{figure}[t]
\begin{center}
\includegraphics[width=0.30\textwidth,clip]{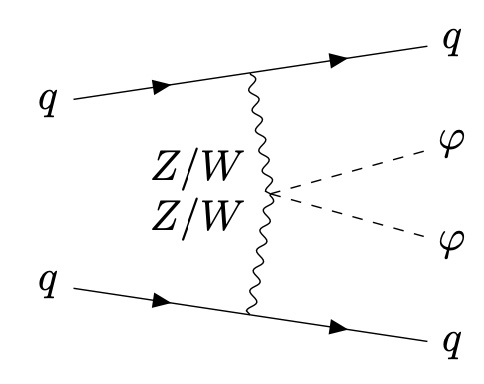}	\qquad\quad
\includegraphics[width=0.30\textwidth,clip]{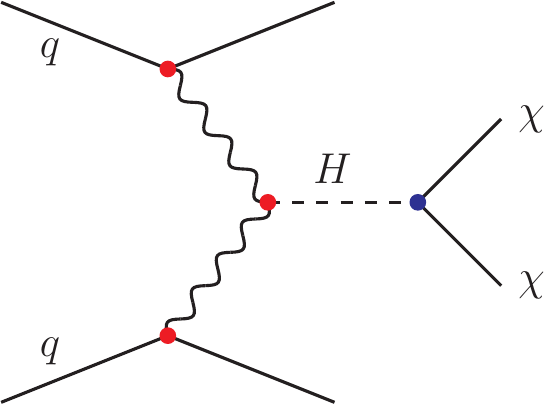}
\end{center}
\caption{(Left) Feynman diagram for the dark matter production.  (Right) Feynman diagram for a signal process with an invisible decay of additional scalar boson. Adapted from Fig.~1 in~\cite{ATLAS:2022yvh}.
Here, $\chi$ corresponds to DM or an invisible particle and $H$ is an additional scalar boson.}
\label{fig:diagram_signal}
\end{figure}

On the other hand, the signal process for DM production in our work is based on
\begin{align}
pp \to \varphi \varphi j j,
\label{eq:process_sig}
\end{align}
where $j = u, d, c, s, b$ (and their antiparticles).
The Feynman diagram for main signal processes is shown in Fig.~\ref{fig:diagram_signal}.

The main background process is 
\begin{align}
pp \to \nu \bar{\nu} j j,
\label{eq:process_bg}
\end{align}
where $\nu$ $(\bar{\nu} )$ is (anti-) neutrino and is summed over three flavors 
and $j$ not only includes $u, d, c, s, b$ (and their antiparticles) but also gluons.
Because there is no interference between the SM background process~\eqref{eq:process_bg} and the signal process~\eqref{eq:process_sig},
the cross section is suppressed by $C^2/\Lambda^8$ schematically, with $C$ being the Wilson coefficient.

Before discussing the HL-LHC search, we briefly mention the current ATLAS measurement
with $139$~$\mathrm{fb}^{-1}$ of LHC $p p$ collision data at a center-of-mass energy of 
$\sqrt{s} = 13$~TeV recorded by ATLAS detector~\cite{ATLAS:2022yvh}.
They considered a signal process of \eqref{eq:process_sig}, but a scalar boson was introduced as a mediator to the DM, as shown on the right plot of Fig.~\ref{fig:diagram_signal}.

The dimension-8 effective operators in our work are regarded as a consequence of integrating out heavy resonances. 
For a heavy scalar mass of $1$~TeV, we take the bounds on dark matter production from the results of Fig.~14 in Ref~\cite{ATLAS:2022yvh}.
As shown Table~\ref{tab:atlas_recast}, we set $\Lambda = 1$~TeV and the DM mass $m_{\varphi} = 375$~GeV $\sim 3m_h$, and show that the Wilson coefficients above $C^{(1)}_{H^2\varphi^2} = C^{(2)}_{H^2\varphi^2} = 32$
are excluded by the current LHC data.
When either of coefficients is turned off, $C^{(2)}_{H^2\varphi^2}$ can undergo more severe constraints than $C^{(1)}_{H^2\varphi^2}$.

\begin{table}[hbt!]
\center
\begin{tabular}{|c|c|}
\hline
$\sqrt{s} = 13$~TeV LHC, $L_{\mathrm{int}} = 139$~$\mathrm{fb}^{-1}$ & $\sigma^{\mathrm{VBF}}\times B_{\mathrm{inv}} = 0.11$~pb ($m_H = 1$~TeV)\\
\hline\hline
$\Lambda = 1$~TeV, $m_{\varphi} = 375$~GeV & cross section from EFT operators \\\hline
$(C^{(1)}_{H^2\varphi^2}, C^{(2)}_{H^2\varphi^2}) = (40, 40)$ & $0.28$~pb\\\hline
$(C^{(1)}_{H^2\varphi^2}, C^{(2)}_{H^2\varphi^2}) = (32, 32)$ & $0.11$~pb\\\hline
$(C^{(1)}_{H^2\varphi^2}, C^{(2)}_{H^2\varphi^2}) = (40, 0)$ & $0.012$~pb\\\hline
$(C^{(1)}_{H^2\varphi^2}, C^{(2)}_{H^2\varphi^2}) = (0, 40)$ & $0.097$~pb\\
\hline
\end{tabular}
\caption{Comparison between a cross section at $95$~\% CL upper limit in Fig.~14 in~\cite{ATLAS:2022yvh} (first line) and cross sections from the dimension-8 Higgs portal operators (3rd--6th lines). 
In the first line, $m_H$ is the mass of a heavy scalar mediator. The mediator decays invisibly with a branching ratio $B_{\mathrm{inv}}$.
In the 3rd--6th lines, the DM mass is fixed with $375$~GeV $\sim 3m_h$ and $\Lambda = 1$~TeV.}
\label{tab:atlas_recast}
\end{table}

We can translate the mass of a new heavy resonance of $M \geq 1$~TeV to the Wilson coefficients of the dimension-8 operators~\cite{Boughezal:2022nof} by
\begin{align}
\frac{\Lambda}{(|C|)^{1/4}} \geq \frac{1~\mathrm{TeV}}{\sqrt{g}},
\end{align}
where $C$ is the Wilson coefficient of a dimension-8 operator and $g$ is the coupling of the heavy resonance.
Then, if we take $g = \sqrt{4\pi}$ at maximum and $\Lambda = 1$~TeV,
we have $|C| \leq 13$.
Thus, $C = 40$ corresponds to $\Lambda \sim 400$~GeV $(\sim 1~\mathrm{TeV}/\sqrt{6.3})$ for the normalization $|C| = 1$, which 
is slightly smaller than $1~\mathrm{TeV}/\sqrt{g}\sim 530$~GeV for $g = \sqrt{4\pi} \sim 3.5$, so it might be acceptable to scan up to $|C| \leq 40$ in our EFT analysis.

For the HL-LHC search, we can benefit from different features of a signal from each operator, $O^{(1)}_{H^2\varphi^2}$ or  $O^{(2)}_{H^2\varphi^2}$.
They may show different kinematical distributions similarly to Refs.~\cite{Alioli:2020kez,Li:2022rag,Boughezal:2022nof}.
The scattering processes, $W^+ W^- \to \varphi \varphi$ and  $Z Z \to \varphi \varphi$, 
shows different dependencies on the Mandelstam variables, depending on the operators. 
Namely, with $O^{(2)}_{H^2\varphi^2}$ only, we have  the Mandelstam variable $s$ and mass dependencies, whereas
there is an additional  dependency on the Mandelstam variable $t$ in the presence of $O^{(1)}_{H^2\varphi^2}$.


\section{Conclusions}\label{sec:summary}

We have presented the positivity bounds on the dimension-8 Higgs-portal interactions for WIMP scalar dark matter by taking the superposed states for Higgs and scalar dark matter. From the results, we showed that the positivity bounds can curb out the part of the parameter space for the effective couplings, otherwise unconstrained by phenomenological consideration, such as dark matter relic density, direct and indirect detection and LHC constraints. 

Motivated by the effective theory coming from massive graviton or radion, we worked closely to the parameter space where the universal condition for other effective Higgs-portal couplings is imposed and the strong bounds from direct detection experiments such as the LZ experiment can be avoided. 
Even in this case, we showed that there are interesting signatures for cosmic ray observations due to the dark matter annihilations into a pair of heavy gauge bosons such as a pair of Higgs bosons, $WW$ or $ZZ$.

\def\theequation{A.\arabic{equation}}

\setcounter{equation}{0}

\vskip0.8cm
\noindent
{\Large \bf Appendix: One-loop corrections to the positivity bounds}

\vspace{0.5cm}

In this appendix, we consider the one-loop corrections to the Wilson coefficients for the Higgs-portal dimension-8 operators in the presence of dimension-6 operators in the effective theory.

We first consider the dimension-6 Higgs-portal terms with two derivatives introduced in the text, in the following,
\bea
{\cal L}_{\rm dim-6} =\frac{1}{3\Lambda^4} \Big({\tilde d}_3\varphi^2 |D_\mu H|^2 + {\tilde d}_4 |H|^2 (\partial_\mu\varphi)^2\Big) \label{dim6}
\eea
with ${\tilde d}_3\equiv d_3 m^2_\varphi, {\tilde d}_4\equiv d_4 m^2_H$.
So, the Feynman rule for the Higgs-portal four-point vertex is given by
\bea
[\varphi(k),\varphi(k'),H(p),H^\dagger(p')] _{\rm dim-6}= \frac{2i}{3\Lambda^4} \Big({\tilde d}_3 (p'\cdot p)-{\tilde d}_4(k\cdot k') \Big)
\eea
where $k, k', p$ are the incoming momenta into the vertex and $p'$ is the outgoing momentum from the vertex.

On the other hand, we also list the dimension-8 Higgs-portal terms with four derivatives, as follows,
\bea
{\cal L}_{\rm dim-8} =\frac{ C^{(1)}_{H^2\varphi^2}}{\Lambda^4} (D_\mu H)^\dagger (D_\nu H) \partial^\mu\varphi \partial^\nu\varphi +\frac{ C^{(2)}_{H^2\varphi^2}}{\Lambda^4} (D_\mu H)^\dagger (D^\mu H) \partial_\nu\varphi \partial^\nu\varphi.  \label{dim8}
\eea
Then, the corresponding Feynman rule is 
\bea
[\varphi(k),\varphi(k'),H(p),H^\dagger(p')] _{\rm dim-8}&=& -\frac{i  C^{(1)}_{H^2\varphi^2}}{\Lambda^4} \Big((p\cdot k')(p'\cdot k)+(p\cdot k)(p'\cdot k')\Big) \nonumber \\
&& -\frac{2i  C^{(2)}_{H^2\varphi^2}}{\Lambda^4}\,(p\cdot p')(k\cdot k') \nonumber \\
&\to &  \frac{i  C^{(1)}_{H^2\varphi^2}}{4\Lambda^4}(u^2+s^2)+ \frac{i  C^{(2)}_{H^2\varphi^2}}{2\Lambda^4}\,t^2.
\eea
Here, in the second arrow, we took the massless limit for which $s=(p+k)^2= 2p\cdot k =-2p'\cdot k'$, $t=(p-p')^2=-2p\cdot p'=2k\cdot k'$ and $u=(p+k')^2=2p\cdot k'=-2p'\cdot k$, and the energy conservation, $p+k=p'-k'$.

The dimension-6 operators in Eq.~(\ref{dim6}) give rise to one-loop corrections to the Higgs-portal four-point vertex, as follows,
\bea
{\cal M} ={\cal M}_1 +{\cal M}_2
\eea
where
\bea
{\cal M}_1 =  \bigg(\frac{i}{3\Lambda^4}\bigg)^2\int \frac{d^4 q}{(2\pi)^4} \bigg[\frac{i}{q^2-m^2_\varphi} \bigg] \bigg[\frac{i}{(p+k-q)^2-m^2_H}\bigg] \,{\cal N}
\eea
with
\bea
 {\cal N}\equiv 4\Big({\tilde d}_3 p\cdot (p+k-q)+{\tilde d}_4 (k\cdot q) \Big) \Big({\tilde d}_3 p'\cdot (p+k-q)-{\tilde d}_4 (k'\cdot q) \Big), \label{N1}
\eea
and ${\cal M}_2={\cal M}_1(k\leftrightarrow k')$. 

Then, using the momentum conservation, $p+k=p'-k'$, we can rewrite the numerator in Eq.~(\ref{N1}) as
\bea
 {\cal N}&=& 4\Big({\tilde d}_3 p\cdot (p+k-q)+{\tilde d}_4 (k\cdot q) \Big) \Big({\tilde d}_3 p'\cdot (p'-k'-q)-{\tilde d}_4 (k'\cdot q) \Big) \nonumber \\
 &\to &  4\Big({\tilde d}_3 p\cdot (k-q)+{\tilde d}_4 (k\cdot q) \Big) \Big(-{\tilde d}_3  p'\cdot (k'+q)-{\tilde d}_4 (k'\cdot q) \Big).
\eea
Here, we took the massless limit for the external states in the arrow, because we are interested in the corrections to the dimension-8 operators.
 
We first recast the loop momentum integral in the Feynman parametrization and with a shift in the loop momentum by $l=q-x(p+k)$, as follows,
\bea
{\cal M}_1 =\frac{1}{9\Lambda^8} \int^1_0 dx \int \frac{d^4 l}{(2\pi)^4} \frac{{\cal  N}(q=l+x(p+k))}{(l^2-\Delta)^2}
\eea 
with
\bea
\Delta = -x(1-x)(p+k)^2 +(1-x)m^2_\varphi+xm^2_H.
\eea
We note that
\bea
p\cdot q &=& p\cdot l +xp\cdot (p+k) \to p\cdot l + x p\cdot k, \\
k\cdot q &=& k\cdot l + x k\cdot (p+k) \to k\cdot l + x k\cdot p, \\
p'\cdot q &=& p'\cdot l + x p' \cdot (p+k)=p'\cdot l + x p' \cdot (p'-k') \to p'\cdot l - xp'\cdot k', \\
k'\cdot q &=& k'\cdot l + xk'\cdot (p+k) = k'\cdot l +xk'\cdot (p'-k') \to k'\cdot l+ xk'\cdot p',
\eea
where we dropped the mass terms after the arrows. 
Then, we get
\bea
{\cal N}&=&4\bigg(\Big((1-x){\tilde d}_3+x {\tilde d}_4\Big)(p\cdot k)-({\tilde d}_3 p-{\tilde d}_4 k)\cdot l\bigg) \nonumber \\
&&\times \bigg(\Big(-(1-x){\tilde d}_3-x {\tilde d}_4\Big)(p'\cdot k')-({\tilde d}_3 p'+{\tilde d}_4 k')\cdot l\bigg) \nonumber \\
&\to&-4\Big((1-x){\tilde d}_3+x {\tilde d}_4\Big)^2 (p\cdot k)(p'\cdot k') \nonumber \\
&&+4\Big(({\tilde d}_3 p-{\tilde d}_4 k)\cdot l\Big)\Big(({\tilde d}_3 p'+{\tilde d}_4 k')\cdot l\Big).
\eea
Using the non-vanishing loop momentum integrals in dimensional regularization with $d=4-\epsilon$,
\bea
&& \mu^{4-d}\int \frac{d^dl}{(2\pi)^d} \frac{1}{(l^2-\Delta)^2} =\frac{i}{(4\pi)^2} \bigg(\frac{2}{\epsilon}-\gamma+\ln 4\pi-\ln\frac{\Delta}{\mu^2}\bigg), \\
&& \mu^{4-d} \int \frac{d^dl}{(2\pi)^d}\frac{l^\mu l^\nu}{(l^2-\Delta)^2} =\frac{i  g^{\mu\nu}}{2(4\pi)^2} \Delta \bigg(\frac{2}{\epsilon}-\gamma+\ln 4\pi-\ln\frac{\Delta}{\mu^2}\bigg),
\eea
we obtain
\bea
{\cal M}_1 &=&\frac{1}{9\Lambda^8}\frac{i}{(4\pi)^2}  \int^1_0 dx \bigg(\frac{2}{\epsilon}-\gamma+\ln 4\pi-\ln\frac{\Delta}{\mu^2} \bigg) \nonumber \\
&&\times \bigg[-4\Big((1-x){\tilde d}_3+x {\tilde d}_4\Big)^2 (p\cdot k)(p'\cdot k') \nonumber \\
&&\quad +2 \Delta({\tilde d}_3 p-{\tilde d}_4 k)\cdot({\tilde d}_3 p'+{\tilde d}_4 k')  \bigg]. \label{res1}
\eea
Similarly, we also get 
\bea
{\cal M}_2 &=& {\cal M}_1(k\leftrightarrow k') \nonumber \\
&=&\frac{1}{9\Lambda^8}\frac{i}{(4\pi)^2}  \int^1_0 dx \bigg(\frac{2}{\epsilon}-\gamma+\ln 4\pi-\ln\frac{\Delta(k\to k')}{\mu^2} \bigg) \nonumber \\
&&\times \bigg[-4\Big((1-x){\tilde d}_3+x {\tilde d}_4\Big)^2 (p\cdot k')(p'\cdot k) \nonumber \\
&&\quad+2\Delta(k\to k')({\tilde d}_3 p-{\tilde d}_4 k')\cdot({\tilde d}_3 p'+{\tilde d}_4 k)  \bigg]. \label{res2}
\eea

In order to write the loop corrections in the form of dimension-8 operators, we take
\bea
-4(p\cdot k)(p'\cdot k') \to s^2, \quad -4(p\cdot k')(p'\cdot k) \to u^2,
\eea
and
\bea
2\Delta({\tilde d}_3 p-{\tilde d}_4 k)\cdot({\tilde d}_3 p'+{\tilde d}_4 k')\to -2x(1-x)A
\eea
with
\bea
A&=& (p+k)^2 ({\tilde d}_3 p-{\tilde d}_4 k)\cdot({\tilde d}_3 p'+{\tilde d}_4 k') \nonumber \\
&=&(p+k)^2\Big( {\tilde d}^2_3 (p\cdot p')-{\tilde d}^2_4 (k\cdot k')-{\tilde d}_3 {\tilde d}_4 (k\cdot p'-k'\cdot p) \Big) \nonumber \\
&\to &-\frac{1}{2}  ({\tilde d}^2_3+{\tilde d}^2_4) st +{\tilde d}_3 {\tilde d}_4 su.
\label{Aterms}
\eea
Similarly, we get
\bea
2\Delta(k\to k')({\tilde d}_3 p-{\tilde d}_4 k')\cdot({\tilde d}_3 p'+{\tilde d}_4 k)\to -2x(1-x) A'
\eea
with
\bea
A'&=& (p+k')^2({\tilde d}_3 p-{\tilde d}_4 k')\cdot({\tilde d}_3 p'+{\tilde d}_4 k) \nonumber \\
&=&(p+k')^2\Big( {\tilde d}^2_3 (p\cdot p')-{\tilde d}^2_4 (k\cdot k') 
-{\tilde d}_3 {\tilde d}_4 (k'\cdot p' -k\cdot p)\Big)   \nonumber \\
&\to & -\frac{1}{2}({\tilde d}^2_3+{\tilde d}^2_4) ut + {\tilde d}_3 {\tilde d}_4  su.
 \label{Aterms}
\eea
Here, we ignored the mass terms in $\Delta$.

As a result, from Eqs.~(\ref{res1}) and (\ref{res2}), the one-loop corrections to the dimension-8 operators are identified, as follows,
\bea
{\cal M}_1 &=&\frac{1}{9\Lambda^8}\frac{i}{(4\pi)^2}  \int^1_0 dx \bigg(\frac{2}{\epsilon}-\gamma+\ln 4\pi-\ln\frac{[(1-x)m^2_\varphi+xm^2_H-x(1-x)s]}{\mu^2} \bigg)  \nonumber \\
&&\times\bigg[ \Big((1-x){\tilde d}_3+x {\tilde d}_4\Big)^2 s^2+x(1-x)\Big(  ({\tilde d}^2_3+{\tilde d}^2_4) st -2{\tilde d}_3 {\tilde d}_4 su\Big) \bigg],
\eea
and 
\bea
{\cal M}_2 &=&\frac{1}{9\Lambda^8}\frac{i}{(4\pi)^2}  \int^1_0 dx \bigg(\frac{2}{\epsilon}-\gamma+\ln 4\pi-\ln\frac{[(1-x)m^2_\varphi+xm^2_H-x(1-x)u]}{\mu^2} \bigg)  \nonumber \\
&&\times\bigg[ \Big((1-x){\tilde d}_3+x {\tilde d}_4\Big)^2 u^2+x(1-x)\Big(  ({\tilde d}^2_3+{\tilde d}^2_4) ut -2{\tilde d}_3 {\tilde d}_4 su\Big) \bigg],
\eea

Taking $s\simeq u$ to be space-like and $|s|\gg m^2_\varphi, m^2_H$ and performing the integral for the Feynman parameter, we obtain
\bea
{\cal M}_1 &=&\frac{1}{9\Lambda^8}\frac{i}{(4\pi)^2} \bigg(\frac{2}{\epsilon}-\gamma+\ln 4\pi+\ln \frac{\mu^2}{|s|} \bigg) \nonumber \\
&&\times \bigg[\frac{1}{3}\Big( {\tilde d}^2_3 + {\tilde d}^2_4 +{\tilde d}_3 {\tilde d}_4\Big)s^2 +\frac{1}{6} \Big(({\tilde d}^2_3 + {\tilde d}^2_4) st- 2{\tilde d}_3 {\tilde d}_4 su\Big) \bigg] \nonumber \\
&&+\frac{1}{9\Lambda^8}\frac{i}{(4\pi)^2}  \bigg[\Big(\frac{13}{18}( {\tilde d}^2_3 + {\tilde d}^2_4 )+\frac{5}{9} {\tilde d}_3 {\tilde d}_4\Big)s^2+\frac{5}{18} \Big(({\tilde d}^2_3 + {\tilde d}^2_4)st - 2{\tilde d}_3 {\tilde d}_4 su\Big)  \bigg], \label{m1}
\eea
and
\bea
{\cal M}_2 &=&\frac{1}{9\Lambda^8}\frac{i}{(4\pi)^2} \bigg(\frac{2}{\epsilon}-\gamma+\ln 4\pi+\ln \frac{\mu^2}{|s|} \bigg) \nonumber \\
&&\times \bigg[\frac{1}{3}\Big( {\tilde d}^2_3 + {\tilde d}^2_4 -{\tilde d}_3 {\tilde d}_4\Big)u^2 +\frac{1}{6} \Big(({\tilde d}^2_3 + {\tilde d}^2_4)ut - 2{\tilde d}_3 {\tilde d}_4 su\Big) \bigg] \nonumber \\
&&+\frac{1}{9\Lambda^8}\frac{i}{(4\pi)^2}  \bigg[\Big(\frac{13}{18}( {\tilde d}^2_3 + {\tilde d}^2_4 )+\frac{5}{9} {\tilde d}_3 {\tilde d}_4\Big)u^2+\frac{5}{18} \Big(({\tilde d}^2_3 + {\tilde d}^2_4) ut- 2{\tilde d}_3 {\tilde d}_4 su\Big)  \bigg]. \label{m2}
\eea
Summing Eqs.~(\ref{m1}) and (\ref{m2}), we obtain the full one-loop corrections as
\bea
{\cal M} &=& \frac{1}{9\Lambda^8}\frac{i}{(4\pi)^2} \bigg(\frac{2}{\epsilon}-\gamma+\ln 4\pi+\ln \frac{\mu^2}{|s|} \bigg) \nonumber \\
&&\times \bigg[\frac{1}{3}\Big( {\tilde d}^2_3 + {\tilde d}^2_4 +{\tilde d}_3 {\tilde d}_4\Big)(s^2+u^2) +\frac{1}{6} \Big(({\tilde d}^2_3 + {\tilde d}^2_4)(s+u)t - 4{\tilde d}_3 {\tilde d}_4 su\Big) \bigg] \nonumber \\
&&+\frac{1}{9\Lambda^8}\frac{i}{(4\pi)^2}  \bigg[\Big(\frac{13}{18}( {\tilde d}^2_3 + {\tilde d}^2_4 )+\frac{5}{9} {\tilde d}_3 {\tilde d}_4\Big)(s^2+u^2)+\frac{5}{18} \Big(({\tilde d}^2_3 + {\tilde d}^2_4) (s+u)t- 4{\tilde d}_3 {\tilde d}_4 su\Big)  \bigg] \nonumber \\
&=& \frac{1}{9\Lambda^8}\frac{i}{(4\pi)^2} \bigg(\frac{2}{\epsilon}-\gamma+\ln 4\pi+\ln \frac{\mu^2}{|s|} \bigg) \nonumber \\
&&\times \bigg[ \frac{1}{3}( {\tilde d}_3 + {\tilde d}_4)^2(s^2+u^2) -\frac{1}{6} ( {\tilde d}_3 + {\tilde d}_4)^2t^2 \bigg] \nonumber \\
&&+\frac{1}{9\Lambda^8}\frac{i}{(4\pi)^2}  \bigg[\Big(\frac{13}{18}( {\tilde d}^2_3 + {\tilde d}^2_4)+\frac{10}{9} {\tilde d}_3 {\tilde d}_4 \Big) (s^2+u^2)-\frac{5}{18} ( {\tilde d}_3 + {\tilde d}_4)^2 t^2 \bigg]. 
\eea
Here, in the second line, we used $s+u=-t$ and $su=\frac{1}{2}(s+u)^2-\frac{1}{2}s^2-\frac{1}{2}u^2=\frac{1}{2}(t^2-s^2-u^2)$.
Therefore, in $\overline{\rm MS}$ scheme for which the divergent terms in combination of $\frac{2}{\epsilon}-\gamma+\ln 4\pi$ are subtracted, the dimension-8 operators in Eq.~(\ref{dim8}) get renormalized at the renormalization scale $\mu$ as
\bea
 {\hat C}^{(1)}_{H^2\varphi^2} &=&  C^{(1)}_{H^2\varphi^2}+\frac{1}{9(4\pi)^2 \Lambda^4}\,\Big(\frac{26}{9} ({\tilde d}^2_3+{\tilde d}^2_4)+\frac{40}{9}  {\tilde d}_3 {\tilde d}_4 \Big) \nonumber \\
 &&+\frac{1}{9(4\pi)^2 \Lambda^4}\,\frac{4}{3} ({\tilde d}_3+{\tilde d}_4)^2\ln \frac{\mu^2}{|s|}, \\
 {\hat C}^{(2)}_{H^2\varphi^2} &=&  C^{(2)}_{H^2\varphi^2}-\frac{1}{9(4\pi)^2 \Lambda^4}\,\frac{5}{9} ({\tilde d}_3 + {\tilde d}_4)^2   \nonumber \\
 &&- \frac{1}{9(4\pi)^2 \Lambda^4}\,\frac{1}{3}({\tilde d}_3 + {\tilde d}_4)^2\ln \frac{\mu^2}{|s|}.
\eea
The above results are quoted in the text for the modified positivity bounds.


\section*{Acknowledgments}

We thank Gauthier Durieux, Michael Trott, Myeonghun Park, and Nicholas L. Rodd, for their valuable comments and discussions. 
The work is supported in part by Basic Science Research Program through the National Research Foundation of Korea (NRF) funded by the Ministry of Education, Science and Technology (NRF-2022R1A2C2003567 and NRF-2021R1A4A2001897). 
The work of KY is supported by Brain Pool program funded by the Ministry of Science and ICT through the National Research Foundation of Korea (NRF-2021H1D3A2A02038697).


\end{document}